\documentclass[12pt,a4paper]{article}

\newcommand{\tw}{$T_2\oplus T_4$}
\newcommand{\non}{non--\tw}

\begin{document}

%\begin{flushright} RHBNC--TH/5/1 \end{flushright}
\title{
Modular Symmetries of Threshold Corrections for Abelian Orbifolds with Discrete
Wilson Lines
}

\author{A.Love  and  S.Todd\\Royal Holloway and Bedford New College,\\ 
University of London, \\ Egham, Surrey TW20 OEX.}
\date{May 1996}

\maketitle
\thispagestyle{empty}

\begin{abstract}
The modular symmetries of string loop threshold corrections for gauge coupling
constants are studied in the presence of discrete Wilson lines for all examples
of abelian orbifolds, where the point group is realised by the action of 
Coxeter elements or generalised Coxeter elements on the root lattices of the 
Lie groups.
\end{abstract}
\clearpage

%body of paper
\setcounter{page}{1}
\parindent=0cm

\section{Introduction }

Orbifold compactifications of string theory \cite{orb1,orb2} 
possess various moduli, which are
background fields corresponding to marginal deformations of the corresponding 
conformal field theory, including radii and angles of the underlying six 
dimensional torus. The spectrum of states for an orbifold is invariant under
certain discrete transformations of the moduli, together with the winding 
numbers and momenta, referred to as modular symmetries \cite{dijk1,dijk2,shap1,
dine1,lau1,kik1,giv1,ler1}.
These modular 
symmetries also appear as symmetries of the string loop threshold corrections
 \cite{kap1,dix1,iba1,iba2,der1,bai1} which are important for the unification 
of gauge coupling constants. In 
this case, it is only the transformations of the moduli associated with fixed 
planes of twisted sectors of the orbifold theory that are relevant, and it is
these particular modular symmetries we shall focus on here. Moreover, the form 
of the threshold corrections, which is constrained by the modular symmetries,
determines the form of the non--perturbative superpotential due to gaugino 
condensation in the hidden sector and the effective potential, the minimization
of which determines the values of the T and U moduli 
\cite{fon1,dix2,cas1,nil1,lou1,lus1,lus2}.

In the absence of Wilson lines, provided all twisted sector fixed planes are 
such that the six-torus  $T^6$ can be decomposed as the direct sum $T^2\oplus
T^4$ with the fixed plane lying in $T^2$, the group of modular symmetries of 
the threshold corrections is a product of PSL(2,Z) factors, one for each of 
the T and U moduli associated with the fixed planes. When such a decomposition
 is not possible ( the non--$T^2\oplus T^4$ case) the group of modular 
symmetries is in general a product of congruence subgroups \cite{may1,bai2,bai3}
of PSL(2,Z).
Wilson lines can also break the PSL(2,Z) modular symmetries \cite{spa1,erl1,
bai4}
or further break \cite{lov1} the congruence subgroups of PSL(2,Z).

It is our purpose here to classify all possible modular symmetry groups
associated with threshold corrections for abelian ($Z_N$ and $Z_N\otimes Z_M$)
Coxeter and generalised Coxeter orbifolds, for abitrary choices of (discrete)
Wilson lines. In section 2, the Coxeter and generalised Coxeter abelian 
orbifolds are classified. In section 3, the modular symmetry groups in the 
presence of Wilson lines are derived in the $T^2\oplus T^4$ cases, in the 
sense discussed above, and in section 4 they are derived for the non-$T^2\oplus
T^4$ cases. Section 5 contains our conclusions.

\section{Coxeter and Generalised Coxeter Orbifolds}

A large class of abelian ($Z_N$ and $Z_N\otimes Z_M$) orbifolds, 
the Coxeter orbifolds, 
can be obtained by taking the underlying lattice of the six--torus to be a
direct sum of Lie group root lattices and constructing the generators of the
point group from Coxeter elements, generalised Coxeter elements or their powers
for the various root lattices. A Coxeter element is the product of all the Weyl
reflections of the root lattice. When the Dynkin diagram possesses an outer
automorphism, we can also make a generalised Coxeter element by using those 
Weyl reflections associated with points  in the Dynkin diagram that are not
permuted by the outer automorphism together with one of the permuted Weyl
reflections and the outer automorphism itself \cite{kat1}.

The only abelian groups that are able to act as automorphisms of a
 6 dimensional lattice and produce exactly N=1 supersymmetry in the four 
dimensional compactified theory \cite{orb1} 
are $Z_N$ and $Z_M\otimes Z_N$ discrete subgroups of 
SU(3). These groups are given in tables 1 and 2 together with the action of
the point group on the 3 complex planes in the space basis (i.e. with the 
bosonic degrees of freedom for the string on the compact manifold $X^1\ldots 
X^6$, written as the combinations ${1\over {\sqrt 2}}(X^1+iX^2)\,, 
{1\over {\sqrt 2}}(X^3+iX^4)$, and ${1\over {\sqrt 2}}(X^5+iX^6).$)
We need all possible choices of Lie group root lattices and Coxeter or 
generalised Coxeter elements which are able to realise these point groups. The 
order N of the Coxeter element is given by \cite{kat1},
\begin{eqnarray}
N={{\rm number\ of\ non-zero\ roots}\over {\rm rank\ of\ Lie\ algebra}}
\end{eqnarray}
In the case of a generalised Coxeter element, the rank of the algebra is
replaced by an effective rank which is the rank reduced by the order of the 
outer automorphism minus one. The possible root lattices of dimensions less
than or equal to 6 where the Coxeters or generalised Coxeters realise the
various relevant $Z_N$ point groups are given in table 3.

Even if the correct
$Z_N$ point group is realised by Coxeters or generalised Coxeters acting on Lie
group root lattices, it is still necessary to check that their eigenvalues
correspond to the correct action of the point group element on the complex
planes in the space basis. This eliminates many possibilities. All possible
choices of lattices for Coxeter and generalised Coxeter $Z_N$ orbifolds have 
already been presented in \cite{kat1}, 
and are included in table 1 for convenience.
For $Z_N\times Z_M$ orbifolds, some examples have already been given 
\cite{kob1,bai5}. If
we insist on realising the point group using only Coxeters and generalised 
Coxeters but not their powers, the only possibilities are as in table 2. Some
$Z_N\times Z_M$ point groups cannot be realised at all in this way.
Appendix B contains the explicit action of the point group on the lattices for
each $Z_N$ or $Z_M\times Z_N$ example. The matrices for the various Coxeter and
generalised Coxeter elements involved are given in appendix A.

In the absence of Wilson lines, the modular symmetries of threshold 
corrections and the threshold corrections themselves have already been 
calculated for all $Z_N$ Coxeter orbifolds \cite{may1,bai2,bai3}
with non--decomposable twisted
sector fixed planes. It remains to carry out the calculations for the 
generalised Coxeter orbifold $Z_6$--II--f with a \non\ $Z_2$ fixed
plane in the $\theta^2$ twisted sector. This is done in appendix C. None of 
the $Z_M\times Z_N$ example have \non\ planes.

\section{Effect of Wilson Lines on Modular Symmetries for Decomposable Fixed 
Planes}

As discussed in the introduction, it is the orbifold twisted sectors that 
possess fixed planes that are relevant to threshold corrections to gauge 
coupling constants. In this section, we shall discuss the modular symmetries 
associated with these twisted sectors in the presence of Wilson lines for the
simplest case where the 6 torus $T^6$ can be decomposed as a direct sum
$T^2\oplus T^4$ with the fixed plane lying in the $T^2$.

The case of $Z_N$ Coxeter or generalised Coxeter orbifolds is discussed first.
Let the Wilson line associated with the lattice vector $e^i_a$ be denoted by
$A^I_a$ where \textbf{I} is a space index for the $E_8\times E_8$ degrees of
freedom and \textbf{a} is a lattice index for $T^6$. The possible values for
the Wilson lines are constrained by the usual \cite{fon2} 
homomorphism and worldsheet 
modular invariance conditions. For a $Z_N$ orbifold these are given by,
\begin{eqnarray}
N A^I_a \in \Lambda_{E_8\times E_8}\label{homcon}\label{eqonlat} 
\end{eqnarray}
where $\Lambda_{E_8\times E_8}$ is the $E_8\times E_8$ lattice, and
\begin{eqnarray}
N(\sum_I (nV^I + r_aA^I_a)^2-n^2v^2)=0\,\pmod{2}\end{eqnarray}
where \( n=0,1\ldots N-1\,,\, r_a=0,1,\ldots N-1\,,\, V^I \) is the embedding
of the point group element $\theta$ in $E_8\times E_8$ and the action of 
$\theta$ in the $j$th complex plane is $e^{2\pi iv_j},\,j=1,2,3.$
\begin{eqnarray}
N \sum_I (A^I_a)^2=0\pmod{2} \label{na2} \\ 
N \sum_I A^I_aA^I_b=0\pmod{1}, a\neq b \label{naa} \\
N \sum_I V^IA^I_a=0\pmod{1}. \label{nva}
\end{eqnarray}

If the action of the point group element $\theta$ on the basis vectors of 
the compact manifold is
\begin{eqnarray}
\theta \ : e_a\rightarrow \Theta_{ab} e_b \end{eqnarray}
then
\begin{eqnarray} A_a=\Theta_{ab}A_b+\Lambda_{E_8\times E_8}\label{aath}
\end{eqnarray}
Identifies  equivalent Wilson lines and contains the homomorphism condition 
(\ref{homcon}).
Equivalent Wilson lines are listed in appendix B for the various orbifolds.
Moreover, by subtracting off appropriate lattice vectors we can choose lattice 
vectors in the region
\begin{eqnarray}
\sum_{I=1}^{9} (A^I_a)^2\leq 1,\ \sum_{I=9}^{16} (A^I_a)^2\leq 1\end{eqnarray}

where the two conditions are seen to apply to the components of the Wilson line
in the first and second $E_8$ factors of the gauge group separately.

For the purpose of defining modular symmetries it is more convenient to work
in terms of the matrix $A_{Ab}$ defined by
\begin{eqnarray}
{A^T_{aB}=\sum_I A_a^IE_B^I}\label{atab}\end{eqnarray}
where $E^I_A$ is a basis for the $E_8\times E_8$ lattice. We can write 
(\ref{atab}) as
\begin{eqnarray}
A^T_{aB}=A_a^IX_{IB}^{-1}\end {eqnarray}
where $X$ is the matrix
\begin{eqnarray}
X={1\over 2}\left( 
\matrix{ 2&-2 &0&0&0&0&0&0\cr 0&2&-2&0&0&0&0&0\cr 0&0&0&2&-2&0&0&0 \cr
0&0&0&0&2&-2&0&0\cr 0&0&0&0&0&2&-2&0\cr 0&0&0&0&0&0&2&2\cr
-1&-1&-1&-1&-1&1&1&-1} \right) \end{eqnarray}
In this formulation \cite{spa1,erl1,iba3} 
acceptable Wilson lines must satisfy
\begin{eqnarray}
A(I-Q)\in Z\label{aiq}\end{eqnarray}
and
\begin{eqnarray}
{1\over 2}A^TCA(I-Q)+{1\over 2}(I-Q^*)A^TCA\in Z\label{atcaiq}\end {eqnarray}
where
\begin{eqnarray}
Q=\Theta^T \end{eqnarray}
and
\begin{eqnarray}{Q^*\equiv (Q^{-1})^T}\end{eqnarray}
These conditions reflect the fact that the action of the point group must be
an automorphism of the whole Narain lattice in the presence of Wilson lines 
\cite{iba3}.
Equations (\ref{aiq}),(\ref{atcaiq}) and (\ref{aath}) 
are slightly more restrictive than (\ref{na2}),(\ref{naa}) and (\ref{aath}), 
and in
practice, we find it convenient to generate Wilson lines consistent with
(\ref{na2}),(\ref{naa}) and (\ref{aath}) 
and then retain only those which also satisfy (\ref{aiq}) and (\ref{atcaiq}). 
For a
specific choice of embedding of the point group we would also have to satisfy
(\ref{nva}).

In the case of a $T^2\oplus T^4$ fixed plane, the problem of determining the 
modular symmetries for the moduli associated with this plane reduces to a
2 dimensional problem \cite{bai4}. 
If $T$ and, in the case of a $Z_2$ plane, U, are the 
moduli associated with this fixed plane, then the full PSL(2,Z) group of 
modular transformations may be written as
\begin{eqnarray}
{T\rightarrow {{aT+b}\over{cT+d}},\  a,b,c,d\in Z,\  ad-bc=1}
\label{modTt}\end{eqnarray}
and
\begin{eqnarray}{U\rightarrow {{a'U+b'}\over{c'U+d'}},\  a',b',c',d'\in Z,\  
a'd'-b'c'=1}\label{modTu}\end{eqnarray}
If the fixed plane is a $Z_k$ fixed plane with $k\neq 2$ i.e. if the action
of the point group in that complex plane is $Z_k$ with $k\neq 2$, then only
the $T$ modulus occurs as a variable parameter of the orbifold model.

In general, the Wilson lines associated with this plane break the PSL(2,Z)
modular groups to subgroups. For the $T$ modular transformations, the 
conditions that determine this subgroup are \cite{bai4}
\begin{eqnarray}
cA\in Z \label{caz}\\ 
{c\over 2} A^TCA\in Z \\ 
cAJA^TC \in Z \\ 
(1-d)A-{c\over 2}AJA^TCA
\in Z \\ 
(1-a)CA -{c\over 2}CAJA^TCA\in Z \\ 
(1-{a\over 2}-{d\over 2})A^TCA-
{c\over 4}A^TCAJA^TCA \in Z
\end{eqnarray}
where $C$ is the Cartan metric for $E_8\times E_8$ and 
\begin{eqnarray}{J=\left( \matrix{ 0&1\cr -1&0\cr}\right).}\label{Jmatrix}
\end{eqnarray}
For the U modular transformations, the conditions that determine the unbroken 
subgroup are
\begin{eqnarray}
A(I-F^{-1})\in Z \\ A^TCA-{1\over 2}A^TCAF^{-1}-{1\over 2}F^TA^TCA\in Z
\label{Fdb}
\end{eqnarray}
where
\begin{eqnarray}{ F=\left( \matrix{d'&b'\cr c'&a'}\right)}\end{eqnarray}
In (\ref{caz})--(\ref{Fdb}) 
the fixed plane has been treated as an independent 2 dimensional plane
and $A$ and $A^T$ have been truncated to the relevant two columns and rows 
respectively, which correspond to the fixed plane.

We have determined all possible modular symmetry groups in the presence of 
Wilson lines for the \tw\ fixed planes of $Z_N$ Coxeter and generalised
Coxeter orbifolds. These are given in appendix D along with examples of
Wilson lines that break the full PSL(2,Z) group to these subgroups. For the case
of a \tw\ plane the resulting modular groups do not depend directly on
the specific $Z_N$ orbifold, but only on the action of the point group in 
that fixed plane. However, if there are Wilson lines present in other complex
planes they can affect the allowed values of the Wilson lines in the fixed plane
through the world sheet modular invariance conditions or (\ref{atcaiq}), 
because the Cartan
matrix $C$ links rows of $A^T$ corresponding to different complex planes. In 
that case, the full action of the point group is of importance because of its
influence on inequivalent Wilson lines in the other complex planes.
We have presented all modular groups permitted for the fixed plane with
non--zero Wilson lines only in the fixed plane. With non--zero Wilson lines also
present in other complex planes the possibilities are reduced because of the
restriction of the 
choices of Wilson lines in the fixed plane just discussed. Specific choices 
of the point group embedding will further limit the choices of Wilson lines 
and so the possible modular groups.

For the $Z_M\times Z_N$ orbifolds, the allowed Wilson lines are determined 
by equivalence conditions associated with both point group elemnents
$\theta$ and $\omega$ (as listed in appendix B)  and by the worldsheet modular
invariance conditions \cite{fon2}
defined for the $\theta^k\omega^l$ twisted sector
\begin{eqnarray}{N'\left( \sum_I (kA^I+lB^I+r_aA_a^I)^2 - (ka+lb)^2\right) =0
\pmod{2} }\end{eqnarray}
where $k=0,1,\ldots ,M-1,\  l=0,1,\ldots ,N-1\,,\,N'$ is the order of
$\theta^k\omega^l$, $A^I$ and $B^I$ are the embeddings of $\theta$ and 
$\omega$ in the $E_8\times E_8$ and the action of $\theta$ and $\omega$
on the $j$th complex plane is $e^{2\pi ia_j},\  j=1,2,3$, and 
$e^{2\pi ib_i},\  i=1,2,3$. The integers $r_a$ take values $0,1,\ldots ,N'-1$.

In general, there could be two sets of conditions of the type (\ref{aiq}) 
and (\ref{atcaiq}) to 
satisfy, one for $Q(\theta )$ and one for $Q(\omega )$.  However, by inspection
of appendix B we see that for the most part, only one of $Q(\theta )=\Theta ^T$
$Q(\omega )=\Omega^T$ acts non--trivially in a particular fixed plane, or 
$Q(\theta )$ and $Q(\omega )$ both act in the same way. The only cases where
$Q(\theta )$ and $Q(\omega )$ act differently are when one acts as $C^T(SU(3))$
and the other acts as $C^T(SU(3)^{[2]})$. However, in these cases there is no
breaking of the PSL(2,Z) modular groups. The resulting modular symmetry groups
in the presence of Wilson lines in the fixed plane are given in appendix E.

\section{Effect of Wilson Lines on Modular Symmetries for \non\ 
Fixed Planes}

When a fixed plane in an orbifold twisted sector is such that the 6 torus $T^6$
can \emph{not} be decomposed into a direct sum $T^2\oplus T^4$ with the fixed
plane lying in the $T^2$, the associated modular symmetry groups for $T$ and
$U$ moduli are, in general, already broken to subgroups of PSL(2,Z) even
in the absence of Wilson lines \cite{may1,bai2,bai3}. 
In  the presence of Wilson lines further
breaking of the symmetry can occur. A systematic method of calculating the
resulting modular symmetry groups has been given elsewhere \cite{lov1}. 
The system of
equations that determine these modular groups depends on two matrices 
\cite{mythesis}, 
which we call R and K. The role of these two matrices is that of allowing the
initially 6 dimensional problem to be reduced to a 2 dimensional form. In 
appendix E we list all the R and K matrices for the models considered and
in appendix F we
explain the derivation of these matrices and thus how they relate to
the notation in previous papers (\cite{lov1} and \cite{bai5}). 
There are no $Z_M\times Z_N$
orbifolds with point groups elements realised by Coxeter or generalised Coxeter
elements with \non\ fixed planes, with the exeception of 
$Z_2\times Z_2$ which was considered in \cite{bai6}.

To solve for the modular symmetries, it is convenient to construct an 
$8\times 2$ matrix, which we write simply as $A_T$ in the following, which 
contains the relevant Wilson line components involved in the problem and is 
obtained in the reduction from the 6 dimensional problem, as detailed in
appendix F. The modular symmetries, then, are given by the constraint equations,
which are for the T modulus
\begin{eqnarray}
cA_TRJK \in Z \\
cA_TRJKA^t_TC \in Z \\
(1-d)A_TR-{1\over 2}cA_TRJKA^t_TCA_TR \in Z \\
cKA^t_TC\in Z\\
{1\over 2}cKA^t_TCA_TR\in Z \\
{1\over 2}cA^t_TCA_TRJK \in Z \\
(1-a-{1\over 2}cA^t_TCARJK)A^t_TC\in Z \\
(1-{1\over 2}a-{1\over 2}d)A^t_TCA_TR-{1\over 4}cA^t_TCA_TRJKA^t_TCA_TR+
bK^{-1}J \in Z \\
\end{eqnarray}
and for the U modulus
\begin{eqnarray}
A_TR(I-M)\in Z \\
K^{-1}M^*K\in Z \\
A^t_TCA_TR-{1\over 2}A^t_TCA_TRM-{1\over 2}K^{-1}M^*KA^t_TCA_TR \in Z
\end{eqnarray}

where J is given by (\ref{Jmatrix}) 
\begin{eqnarray}{M=\left( \matrix{ a' & -b' \cr -c' & d' } \right) = F^{-1} }
\end{eqnarray}
and $M^*$ is defined to be $(M^{-1})^t$. The integer coefficients 
$a,b,c,d$ and $a',b',c',d'$ in the constraint equations are the 
coefficients of the modular transformations (\ref{modTt},\ref{modTu})
on the $T$ and $U$ moduli respectively.

\section{Conclusions}

The moduli dependent corrections to gauge coupling constants in orbifold 
compactifactions are a possible mechanism for reconciling the unification 
scale for the standard model gauge coupling constants with string theory.
We have carried out a systematic study of the possible modular symmetry
groups in the presence of quantised Wilson lines for all Coxeter orbifolds.
Modular symmetry constrains the functional form that moduli dependent 
threshold corrections can take and, in some cases, this constraint is strong
enough to explicitly determine the threshold correction. The results of the 
present paper are a first step in that direction for all Coxeter orbifolds.
The modular symmetry groups obtained are in some cases congruence subgroups and
in other cases more exotic subgroups of PSL(2,Z).

\newpage

% Tables
\begin{table}
\begin{tabular}{|l|c|c|}
\hline
Orbifold &	$\theta$	&	Lattice \\
\hline
$Z_3$	 &	$(1,1,-2)/3$	&	$SU(3)\times SU(3)\times SU(3)$	\\
$Z_4-a$ &	$(1,1,-2)/4$	&	$SU(4)\times SU(4)$ 		\\
$Z_4-b$ &	$(1,1,-2)/4$	&	$SU(4)\times SO(5)\times SU(2)$	\\
$Z_4-c$ &	$(1,1,-2)/4$	&	$SO(5)\times SU(2)\times SO(5)
\times SU(2)$	\\
$Z_6--I--a$&	$(-2,1,1)/6$	&	$SU(3)\times G_2\times G_2$	\\
$Z_6--I--b$&	$(-2,1,1)/6$	&	$SU(3)\times SU(3)^{[2]}\times 
SU(3)^{[2]}$ \\
$Z_6-I-c$&	$(-2,1,1)/6$ 	&	$SU(3)\times SU(3)^{[2]}\times G_2$\\
$Z_6-II-a$&	$(2,1,-3)/6$	&	$SU(6)\times SU(2)$		\\
$Z_6-II-b$&	$(2,1,-3)/6$	&	$SU(3)\times SO(8)$		\\
$Z_6-II-c$&	$(2,1,-3)/6$	&	$SU(3)\times SO(7)\times SU(2)$	\\
$Z_6-II-d$&	$(2,1,-3)/6$	&	$SU(3)\times G_2\times SO(4)$	\\
$Z_6-II-e$&	$(2,1,-3)/6$	&	$SU(3)\times SU(3)^{[2]}\times SO(4)$\\
$Z_6-II-f$&	$(2,1,-3)/6$	&	$SU(3)\times SU(4)^{[2]}\times SU(2)$\\
$Z_6-II-g$&	$(2,1,-3)/6$	&	$SU(3)\times Sp(6)\times SU(2)$	\\
$Z_7$&		$(1,2,-3)/7$	&	$SU(7)$				\\
$Z_8-I-a$&	$(1,-3,2)/8$	&	$SO(9)\times SO(5)$		\\
$Z_8-I-b$&	$(1,-3,2)/8$	&	$SO(8)^{[2]}\times SO(5)$	\\
$Z_8-II-a$&	$(1,3,-4)/8$	&	$SO(10)\times SU(2)$		\\
$Z_8-II-b$&	$(1,3,-4)/8$	&	$SO(9)\times SO(4)$		\\
$Z_8-II-c$&	$(1,3,-4)/8$	&	$SO(8)^{[2]}\times SO(4)$	\\
$Z_{12}-I-a$&	$(1,-5,4)/12$	&	$E_6$				\\
$Z_{12}-I-b$&	$(1,-5,4)/12$	&	$F_4\times SU(3)$		\\
$Z_{12}-I-c$&	$(1,-5,4)/12$	&	$SO(8)^{[3]}\times SU(3)$	\\
$Z_{12}-II-a$&	$(1,5,-6)/12$	&	$F_4\times SO(4)$		\\
$Z_{12}-II-b$&	$(1,5,-6)/12$	&	$SO(8)^{[3]}\times SO(4)$	\\
\hline
\end{tabular}
\caption{Coxeter and generalised Coxeter $Z_N$ orbifolds. For the point group
generator $\theta$ we display $(v_1,v_2,v_3)$ such that the action of $\theta$
in the complex orthogonal basis is $(e^{2\pi iv_1},e^{2\pi iv_2},e^{2\pi i
v_2})$. When the generalised Coxeter element with outer automorphism of
order $p$ acts on the root lattice is written as $G^{[p]}$.}
\end{table}

\begin{table}
\begin{tabular}{|l|c|c|c|}
\hline
Orbifold 	&	$\theta$	&	$\omega$	& Lattice\\
\hline
$Z_2\times Z_2$	&	$(1,0,-1)/2$	&	$(0,1,-1)/2$	&$(SO(4))^3$\\
$Z_3\times Z_3$	&	$(1,0,-1)/3$	&	$(0,1,-1)/3$	&$(SU(3))^3$\\
$Z_2\times Z_4$ &	$(1,0,-1)/2$	&	$(0,1,-1)/4$	&--- \\
$Z_4\times Z_4$ & 	$(1,0,-1)/4$	&	$(0,1,-1)/4$	&$(SO(5))^3$\\
$Z_2\times Z_6$ & 	$(1,0,-1)/2$	&	$(0,1,-1)/6$	&---\\
$Z_2\times Z_6'$ & 	$(1,0,-1)/2$	&	$(0,1,-1)/6$	&---\\
$Z_3\times Z_6-a$&	$(1,0,-1)/3$	&	$(0,1,-1)/6$	&$(SU(3))^3$\\
$Z_3\times Z_6-b$&	$(1,0,-1)/3$	&	$(0,1,-1)/6$	&$SU(3)\times 
G_2\times SU(3)$\\
$Z_6\times Z_6-a$&	$(1,0,-1)/6$	&	$(0,1,-1)/6$	&$(G_2)^3$\\
$Z_6\times Z_6-b$&	$(1,0,-1)/6$	&	$(0,1,-1)/6$	&$(G_2)^2\times 
SU(3)$\\
$Z_6\times Z_6-c$&	$(1,0,-1)/6$	&	$(0,1,-1)/6$	&$G_2\times 
(SU(3))^2$\\
$Z_6\times Z_6-d$&	$(1,0,-1)/6$	&	$(0,1,-1)/6$	&$(SU(3))^3$\\
\hline
\end{tabular}
\caption{Coxeter and generalised Coxeter $Z_M\times Z_N$ orbifolds.
For the point group
generator $\theta$ we display $(v_1,v_2,v_3)$ such that the action of $\theta$
in the complex orthogonal basis is $(e^{2\pi iv_1},e^{2\pi iv_2},e^{2\pi i
v_2})$ and similarly for $\omega$.}
\end{table}

\begin{table}
\begin{tabular}{|c|c|}
\hline
Point Group & Possible Root Lattices \\
\hline
$Z_2$	&	$SU(2), SO(4)$\\
$Z_3$	&	$SU(3)$	\\
$Z_4$	&	$SU(4),SO(5)$\\
$Z_6$	&	$SU(6),\,SO(8),\,SO(7),\,Sp(6),$\\
	&	$G_2,\,SU(3)^{[2]},\,SU(4)^{[2]}$\\
$Z_7$	&	$SU(7)$\\
$Z_8$	&	$SO(10),\,SO(9),\,Sp(8),\,SO(8)^{[2]}$\\
$Z_{12}$&	$SO(13),\,Sp(12),\,F_4,\,E_6,\,SO(8)^{[3]}$\\
\hline
\end{tabular}
\caption{Root lattices of dimension$\leq 6$ where the Coxeter or generalised
Coxeter element generate $Z_N$ point groups. The generalised Coxeter elements
with outer automorphism of order $p$ acintg on the root lattice is written as
$G^{[p]}$.}
\end{table}

%Appendices

\appendix
{\bfseries\LARGE Appendices}
\section{Coxeter and Generalised Coxeter Elements}

For each Lie group root lattice the action of the Coxeter element or 
generalised Coxeter element on the basis vectors $e_a$ of the root lattice is
given below. The action of the Weyl reflection $\sigma_a$ associated with the
basis vector $e_a$ on the basis vector $e_b$ is defined to be
\begin{eqnarray}{\sigma_a(e_b)=e_b-{{2e_a\cdot e_b}\over {e_a\cdot e_a}}e_a}
\end{eqnarray}
The generalised Coxeter elements also contain cyclic permutations of the 
basis vectors

\subsection{ SU(2)}
Dynkin diagram \begin{picture}(25,20)(0,-5)\put(10,0){\circle{10}}
\put(10,10){1}\end{picture}

Coxeter $\theta =\sigma_1$

$\theta e_1=-e_1$

\subsection{ SU(3)}
Dynkin diagram \begin{picture}(50,20)(0,-5)
\put(10,0){\circle{10}}\put(15,0){\line(1,0){20}}
\put(40,0){\circle{10}}\put(10,10){1}\put(40,10){2}
\end{picture}

Coxeter $\theta =\sigma_1\sigma_2$

$\theta e_1=e_2$, $\theta e_2=-e_1-e_2$

Eigenvalues $exp(\pm 2\pi i/3)$

Generalised Coxeter $\theta =\sigma_2 \left( \matrix{ 1& &2\cr 2&&1\cr}\right)$

$\theta e_1=-e_2$, $\theta e_2=e_1+e_2$

Eigenvalues $exp(\pm 2\pi i/6)$

\subsection{ SU(4)}
Dynkin diagram \begin{picture}(80,20)(0,-5)
\put(10,0){\circle{10}}\put(15,0){\line(1,0){20}}
\put(40,0){\circle{10}}\put(45,0){\line(1,0){20}}
\put(70,0){\circle{10}}\put(10,10){1}\put(40,10){2}\put(70,10){3}
\end{picture}

Coxeter $\theta =\sigma_1\sigma_2\sigma_3$

$\theta e_1=e_2$, $\theta e_2=e_3$, $\theta e_3=-e_1-e_2-e_3$

Eigenvalues $-1,\pm i$

Generalised Coxeter $\theta =\sigma_1\sigma_2 
\left( \matrix{ 1& &3\cr 3&&1\cr}\right)$

$\theta e_1=e_1+e_2+e_3$, $\theta e_2=-e_1-e_2$, $\theta e_3=e_2$

Eigenvalues $-1,exp(\pm 2\pi i/6)$

\subsection{ SU(6)}
Dynkin diagram  \begin{picture}(110,10)(0,-5)
\put(10,0){\circle{10}}\put(15,0){\line(1,0){20}}
\put(40,0){\circle{10}}\put(45,0){\line(1,0){20}}
\put(70,0){\circle{10}}\put(75,0){\line(1,0){20}}
\put(100,0){\circle{10}}
\put(10,10){1}\put(40,10){2}\put(70,10){3}\put(100,10){4}
\end{picture}

Coxeter $\theta =\sigma_1\sigma_2\sigma_3\sigma_4\sigma_5$

$\theta e_1=e_2$, $\theta e_2=e_3$, $\theta e_3=e_4$, $\theta e_4=e_5$,  
$\theta e_5=-e_1-e_2-e_3-e_4-e_5$

Eigenvalues $-1,exp(\pm 2\pi i/3),exp(\pm 2\pi i/6)$

\subsection{ SU(7)}
Dynkin diagram \begin{picture}(140,20)(0,-5)
\put(10,0){\circle{10}}\put(15,0){\line(1,0){20}}
\put(40,0){\circle{10}}\put(45,0){\line(1,0){20}}
\put(70,0){\circle{10}}\put(75,0){\line(1,0){20}}
\put(100,0){\circle{10}}\put(105,0){\line(1,0){20}}
\put(130,0){\circle{10}}
\put(10,10){1}\put(40,10){2}\put(70,10){3}\put(100,10){4}\put(130,10){5}
\end{picture}

Coxeter $\theta =\sigma_1\sigma_2\sigma_3\sigma_4\sigma_5\sigma_6$

$\theta e_1=e_2$, $\theta e_2=e_3$, $\theta e_3=e_4$, $\theta e_4=e_5$,  
$\theta e_5=e_6$,

$\theta e_6=-e_1-e_2-e_3-e_4-e_5-e_6$

Eigenvalues $exp(\pm 2\pi i/7),exp(\pm 4\pi i/7),exp(\pm 6\pi i/7)$

\subsection{ SO(5)}
Dynkin diagram \begin{picture}(50,20)(0,-5)
\put(10,0){\circle{10}}\put(15,2){\line(1,0){20}}\put(15,-2){\line(1,0){20}}
\put(40,0){\circle*{10}}
\put(10,10){1}\put(40,10){2}
\end{picture}

Coxeter $\theta =\sigma_1\sigma_2$

$\theta e_1=e_1+2e_2$, $\theta e_2=-e_1-e_2$

Eigenvalues $\pm i$

\subsection{ SO(7)}
Dynkin diagram \begin{picture}(80,20)(0,-5)
\put(10,0){\circle{10}}\put(15,0){\line(1,0){20}}
\put(40,0){\circle{10}}\put(45,2){\line(1,0){20}}\put(45,-2){\line(1,0){20}}
\put(70,0){\circle*{10}}
\put(10,10){1}\put(40,10){2}\put(70,10){3}
\end{picture}

Coxeter $\theta =\sigma_1\sigma_2\sigma_3$

$\theta e_1=e_2$, $\theta e_2=e_1+e_2+2e_3$, $\theta e_3=-e_1-e_2-e_3$

Eigenvalues $-1,exp(\pm 2\pi i/6)$

\subsection{ SO(8)}
Dynkin diagram \begin{picture}(90,60)(0,-30)
\put(10,0){\circle{10}}\put(15,0){\line(1,0){20}}
\put(40,0){\circle{10}}\put(45,0){\line(1,1){20}}\put(69,23){\circle{10}}
\put(45,0){\line(1,-1){20}}\put(69,-23){\circle{10}}
\put(10,10){1}\put(40,10){2}\put(79,23){3}\put(79,-23){4}
\end{picture}

Coxeter $\theta =\sigma_1\sigma_2\sigma_3\sigma_4$

$\theta e_1=e_2$, $\theta e_2=e_1+e_2+e_3+e_4$, $\theta e_3=-e_1-e_2-e_3$,
$\theta e_4=-e_1-e_2-e_4$

Eigenvalues $-1$ (\emph{twice}),\ $exp(\pm 2\pi i/6)$

First generalised Coxeter $\theta =\sigma_1\sigma_2\sigma_3 
\left( \matrix{ 3& &4\cr 4&&3\cr}\right)$

$\theta e_1=e_2$, $\theta e_2=e_3$, $\theta e_3=e_1+e_2+e_4$, 
$\theta e_4=-e_1-e_2-e_3$,

Eigenvalues $exp(\pm 2\pi i/8),exp(\pm 10\pi i/8)$

Second generalised Coxeter $\theta =\sigma_1\sigma_2
\left( \matrix{1& 3& 4\cr 3&4&1\cr}\right)$

$\theta e_1=e_1+e_2+e_3$, $\theta e_2=-e_1-e_2$, $\theta e_3=e_1+e_2+e_4$, 
$\theta e_4=e_2$

Eigenvalues $exp(\pm 2\pi i/12),exp(\pm 7\pi i/6)$

\subsection{ SO(9)}
Dynkin diagram  \begin{picture}(110,20)(0,-5)
\put(10,0){\circle{10}}\put(15,0){\line(1,0){20}}
\put(40,0){\circle{10}}\put(45,0){\line(1,0){20}}
\put(70,0){\circle{10}}\put(75,-2){\line(1,0){25}}\put(75,2){\line(1,0){25}}
\put(100,0){\circle*{10}}
\put(10,10){1}\put(40,10){2}\put(70,10){3}\put(100,10){4}
\end{picture}

Coxeter $\theta =\sigma_1\sigma_2\sigma_3\sigma_4$

$\theta e_1=e_2$, $\theta e_2=e_3$, $\theta e_3=e_1+e_2+e_3+2e_4$,
$\theta e_4=-e_1-e_2-e_3-e_4$

Eigenvalues $exp(\pm 2\pi i/8),exp(\pm 6\pi i/8)$

\subsection{ SO(10)}
Dynkin diagram  \begin{picture}(120,60)(0,-30)
\put(10,0){\circle{10}}\put(15,0){\line(1,0){20}}
\put(40,0){\circle{10}}\put(45,0){\line(1,0){20}}\put(70,0){\circle{10}}
\put(75,0){\line(1,1){20}}\put(99,23){\circle{10}}
\put(75,0){\line(1,-1){20}}\put(99,-23){\circle{10}}
\put(10,10){1}\put(40,10){2}\put(70,10){3}\put(109,23){4}\put(109,-23){5}
\end{picture}

Coxeter $\theta =\sigma_1\sigma_2\sigma_3\sigma_4\sigma_5$

$\theta e_1=e_2$, $\theta e_2=e_3$, $\theta e_3=e_1+e_2+e_3+e_4+e_5$,
$\theta e_4=-e_1-e_2-e_3-e_4$,  $\theta e_5=-e_1-e_2-e_3-e_5$

Eigenvalues $-1, exp(\pm 2\pi i/8),exp(\pm 6\pi i/8)$

\subsection{ $G_2$}
Dynkin diagram  \begin{picture}(50,10)(0,-5)
\put(10,0){\circle{10}}\put(15,3){\line(1,0){25}}
\put(15,0){\line(1,0){20}}\put(15,-3){\line(1,0){25}}
\put(40,0){\circle*{10}}
\put(10,10){1}\put(40,10){2}
\end{picture}

Coxeter $\theta =\sigma_1\sigma_2$

$\theta e_1=e_1+2e_2$, $\theta e_2=-e_1-e_2$

Eigenvalues $\pm i$

\subsection{ $F_4$}
Dynkin diagram \begin{picture}(80,10)(0,-5)
\put(10,0){\circle{10}}\put(15,2){\line(1,0){20}}\put(15,-2){\line(1,0){20}}
\put(40,0){\circle*{10}}\put(45,0){\line(1,0){20}}
\put(70,0){\circle{10}}
\put(10,10){1}\put(40,10){2}\put(70,10){3}
\end{picture}

Coxeter $\theta =\sigma_1\sigma_2\sigma_3\sigma_4$

$\theta e_1=e_2$, $\theta e_2=e_1+e_2+e_3$, $\theta e_3=e_4$, 
$\theta e_4=-2e_1-2e_2-e_3-e_4$

Eigenvalues $exp(\pm 2\pi i/12),exp(\pm 10\pi i/12)$

\subsection{ $E_6$}
Dynkin diagram  \begin{picture}(140,30)(0,-20)
\put(10,0){\circle{10}}\put(15,0){\line(1,0){20}}
\put(40,0){\circle{10}}\put(45,0){\line(1,0){20}}
\put(70,0){\circle{10}}\put(75,0){\line(1,0){20}}
\put(100,0){\circle{10}}\put(105,0){\line(1,0){20}}
\put(130,0){\circle{10}}\put(70,5){\line(0,1){15}}
\put(70,25){\circle{10}}
\put(7,-17){1}\put(37,-17){2}\put(67,-17){3}\put(97,-17){5}\put(127,-17){6}
\put(67,30){4}
\end{picture}

Coxeter $\theta =\sigma_1\sigma_2\sigma_3\sigma_4\sigma_5\sigma_6$

$\theta e_1=e_2$, $\theta e_2=e_3$, $\theta e_3=e_1+e_2+e_3+e_4+e_5$, 
$\theta e_4=-e_1-e_2-e_3-e_4$,  $\theta e_5=e_6$,
$\theta e_6=-e_1-e_2-e_3-e_5-e_6$,

Eigenvalues $exp(\pm 2\pi i/3),exp(\pm 2\pi i/12),exp(\pm 10\pi i/12)$

\subsection{ Sp(6)}
Dynkin diagram  \begin{picture}(80,20)(0,-5)
\put(10,0){\circle{10}}\put(15,0){\line(1,0){20}}
\put(40,0){\circle*{10}}\put(45,2){\line(1,0){20}}\put(45,-2){\line(1,0){20}}
\put(70,0){\circle{10}}
\put(10,10){1}\put(40,10){2}\put(70,10){3}
\end{picture}

Coxeter $\theta =\sigma_1\sigma_2\sigma_3$

$\theta e_1=e_2$, $\theta e_2=e_1+e_2+e_3$, $\theta e_3=-2e_1-2e_2-e_3$

Eigenvalues $-1,exp(\pm 2\pi i/6)$

\subsection{ Sp(8)}
Dynkin diagram  \begin{picture}(110,20)(0,-5)
\put(10,0){\circle{10}}\put(15,0){\line(1,0){20}}
\put(40,0){\circle{10}}\put(45,0){\line(1,0){20}}
\put(70,0){\circle*{10}}\put(75,-2){\line(1,0){20}}\put(75,2){\line(1,0){20}}
\put(100,0){\circle{10}}
\put(10,10){1}\put(40,10){2}\put(70,10){3}\put(100,10){4}
\end{picture}

Coxeter $\theta =\sigma_1\sigma_2\sigma_3\sigma_4$

$\theta e_1=e_2$, $\theta e_2=e_3$, $\theta e_3=e_1+e_2+e_3+e_4$, 
$\theta e_4=-2e_1-2e_2-2e_3-e_4$

Eigenvalues $exp(\pm 2\pi i/8),exp(\pm 6\pi i/8)$

\section{Coxeter and Generalised Coxeter Orbifolds}

For each orbifold, the action of the point group on the complex space basis
$exp(2\pi i[v_1,v_2,v_3])$ is specified by $(v_1,v_2,v_3)$. The action of the
point group on the lattice basis is given in terms of Coxeter elements or
generalised Coxeter elements for the root lattices involved, with the Coxeter 
element for the root lattice of the Lie group $G$ written as $C(G)$ and any 
generalised Coxeter element written as $C(G^{[p]})$ where $p$ is the order 
of the outer automorphism involved. Then the action in the lattice basis
of the point group element $\theta$ is given in terms of a matrix $\Theta$
composed of Coxeter element matrices, such that the effect of $\theta$ on the
basis vectors $e_a$ of the lattice is $e_a\rightarrow \Theta_{ab}e_b$. The
order of the point group for the various \tw\ and \non\ twsited sector fixed
planes (in the sense of the introduction) is given. Finally, the equivalent 
Wilson line $A_i$ corresponds to the basis vector $e_i$.

\subsection{$Z_3\ \ \ \theta=(1,1,-2)/3\ \ \ (SU(3))^3$}
$\Theta=[C(SU(3)),\ C(SU(3)),\ C(SU(3)]$

\emph{ no fixed planes.}

$A_2\approx A_1,\, A_4\approx A_3,\,A_6\approx A_5,\ 3A_1\approx 3A_2
\approx 3A_5\approx 0$

\subsection{$Z_4-a\ \ \ \theta=(1,1,-2)/4\ \ \ (SU(4))^2$}
$\Theta=[C(SU(4)),\ C(SU(4))]$

\emph{\non\ $Z_2$ fixed plane in the $\theta^2$ twisted sector.}

$A_3\approx A_2\approx A_1,\, A_6\approx A_5\approx A_4,\, 4A_1\approx 4A_4
\approx 0$

\subsection{$Z_4-b\ \ \ \theta=(1,1,-2)/4\ \ \ SU(4)\times SO(5)\times SU(2)$}
$\Theta=[C(SU(4)),\ C(SO(5)),\ C(SU(2))]$

\emph{\non\ $Z_2$ fixed plane in the $\theta^2$ twisted sector.}

$A_3\approx A_2\approx A_1,\, \ 4A_1\approx A_4\approx 2A_5\approx 2A_6
\approx 0$

\subsection{$Z_4-c\ \ \ \theta=(1,1,-2)/4\ \ \ SO(5)\times SU(2) 
\times SO(5)\times SU(2)$}
$\Theta=[C(SO(5)),\ C(SU(2)),\ C(SO(5)),\ C(SU(2))]$

\emph{\tw\ $Z_2$ fixed plane in the $\theta^2$ twisted sector.}

$A_1\approx A_4\approx 2A_2\approx 2A_3\approx 2A_5\approx 2A_6\approx 0$

\subsection{$Z_6-I-a\ \ \ \theta=(-2,1,1)/6\ \ \ SU(3)\times G_2\times G_2$}
$\Theta=[C(SU(3)),\ C(G_2),\ C(G_2)]$

\emph{\tw\ $Z_3$ fixed plane in the $\theta^3$ twisted sector.}

$A_2\approx A_1,\, 3A_1\approx A_3\approx A_4\approx A_5\approx A_6\approx 0$

\subsection{$Z_6-I-b\ \ \ \theta=(-2,1,1)/6\ \ \ SU(3)\times SU(3)^{[2]}
\times SU(3)^{[2]}$}
$\Theta=[C(SU(3)),\ C(SU(3)^{[2]}),\ C(SU(3)^{[2]})]$

\emph{\tw\ $Z_3$ fixed plane in the $\theta^3$ twisted sector.}

$A_2\approx A_1,\, 3A_1\approx A_3\approx A_4\approx A_5\approx A_6\approx 0$

\subsection{$Z_6-I-c\ \ \ \theta=(-2,1,1)/6\ \ \ SU(3)\times SU(3)^{[2]}
\times G_2$}
$\Theta=[C(SU(3)),\ C(SU(3)^{[2]}),\ C(G_2)]$

\emph{\tw\ $Z_3$ fixed plane in the $\theta^3$ twisted sector.}

$A_2\approx A_1,\, 3A_1\approx A_3\approx A_4\approx A_5\approx A_6\approx 0$

\subsection{$Z_6-II-a\ \ \ \theta=(2,1,-3)/6\ \ \ SU(6)\times SU(2)$}
$\Theta=[C(SU(6)),\ C(SU(2))]$

\emph{\non\ $Z_3$ fixed plane in the $\theta^3$ twisted sector.}

\emph{\non\ $Z_2$ fixed plane in the $\theta^2$ twisted sector.}

$6A_1\approx 2A_6\approx 0,\, A_5\approx A_4\approx A_3\approx A_2\approx A_1$

\subsection{$Z_6-II-b\ \ \ \theta=(2,1,-3)/6\ \ \ SU(3)\times SO(8)$}
$\Theta=[C(SU(3)),\ C(SO(8))]$

\emph{\tw\ $Z_3$ fixed plane in the $\theta^3$ twisted sector.}

\emph{\non\ $Z_2$ fixed plane in the $\theta^2$ twisted sector.}

$A_2\approx A_1,\, 3A_1\approx 0,\  A_3\approx A_4,\ 
2A_3\approx 2A_5\approx 2A_6\approx 0,\, A_3\approx -A_5-A_6$

\subsection{$Z_6-II-c\ \ \ \theta=(2,1,-3)/6\ \ \ SU(3)\times SO(7)\times
 SU(2)$}
$\Theta=[C(SU(3)),\ C(SO(7)),\ C(SU(2))]$

\emph{\tw\ $Z_3$ fixed plane in the $\theta^3$ twisted sector.}

\emph{\non\ $Z_2$ fixed plane in the $\theta^2$ twisted sector.}

$A_2\approx A_1,\, 3A_1\approx A_3\approx A_4\approx 2A_5\approx 2A_6\approx 0$

\subsection{$Z_6-II-d\ \ \ \theta=(2,1,-3)/6\ \ \ SU(3)\times G_2\times
 SO(4)$}
$\Theta=[C(SU(3)),\ C(G_2),\ C(SO(4))]$

\emph{\tw\ $Z_3$ fixed plane in the $\theta^3$ twisted sector.}

\emph{\tw\ $Z_2$ fixed plane in the $\theta^2$ twisted sector.}

$A_2\approx A_1,\, 3A_1\approx A_3\approx A_4\approx 2A_5\approx 2A_6\approx 0$

\subsection{$Z_6-II-e\ \ \ \theta=(2,1,-3)/6\ \ \ SU(3)\times SU(3)^{[2]}
\times SO(4)$}
$\Theta=[C(SU(3)),\ C(SU(3)^{[2]}),\ C(SO(4))]$

\emph{\tw\ $Z_3$ fixed plane in the $\theta^3$ twisted sector.}

\emph{\tw\ $Z_2$ fixed plane in the $\theta^2$ twisted sector.}

$A_2\approx A_1,\, 3A_1\approx A_3\approx A_4\approx 2A_5\approx 2A_6\approx 0$

\subsection{$Z_6-II-f\ \ \ \theta=(2,1,-3)/6\ \ \ SU(3)\times SU(4)^{[2]}
\times SU(2)$}
$\Theta=[C(SU(3)),\ C(SU(4)^{[2]}),\ C(SU(2))]$

\emph{\tw\ $Z_3$ fixed plane in the $\theta^3$ twisted sector.}

\emph{\tw\ $Z_2$ fixed plane in the $\theta^2$ twisted sector.}

$A_2\approx A_1,\, 3A_1\approx A_3\approx 0,\ A_4\approx A_5,\ 
2A_4\approx 2A_6\approx 0$

\subsection{$Z_6-II-g\ \ \ \theta=(2,1,-3)/6\ \ \ SU(3)\times Sp(6)
\times SU(2)$}
$\Theta=[C(SU(3)),\ C(Sp(6)),\ C(SU(2))]$

\emph{\tw\ $Z_3$ fixed plane in the $\theta^3$ twisted sector.}

\emph{\non\ $Z_2$ fixed plane in the $\theta^2$ twisted sector.}

$A_2\approx A_1,\, 3A_1\approx 2A_3\approx 0,\ A_4\approx A_5,\ 
2A_4\approx 2A_6\approx 0$

\subsection{$Z_7\ \ \ \theta=(1,2,-3)/7\ \ \ SU(7)$}
$\Theta=[C(SU(7))]$

\emph{ no fixed planes}

$A_6\approx A_5\approx A_4\approx A_3\approx A_2\approx A_1,\ 7A_1\approx 0$

\subsection{$Z_8-I-a\ \ \ \theta=(1,-3,2)/8\ \ \ SO(9)\times SO(5)$}
$\Theta=[C(SO(9)),\ C(SO(5))]$

\emph{\tw\ $Z_4$ fixed plane in the $\theta^4$ twisted sector.}

$A_1\approx A_2\approx A_3\approx A_5\approx 2A_4\approx 2A_6\approx 0$

\subsection{$Z_8-I-b\ \ \ \theta=(1,-3,2)/8\ \ \ SO(8)^{[2]}\times SO(5)$}
$\Theta=[C(SO(8)^{[2]}),\ C(SO(5))]$

\emph{\tw\ $Z_4$ fixed plane in the $\theta^4$ twisted sector.}

$A_4\approx A_3\approx A_2\approx A_1,\ 2A_1\approx A_5\approx 2A_6\approx 0$

\subsection{$Z_8-II-a\ \ \ \theta=(1,3,-4)/8\ \ \ SO(10)\times SU(2)$}
$\Theta=[C(SO(10)),\ C(SU(2))]$

\emph{\non\ $Z_2$ fixed plane in the $\theta^2$ twisted sector.}

$A_1\approx A_2\approx A_3,\ A_4\approx -A_5,\ 2A_4\approx A_3,\ 2A_3
\approx 2A_6\approx 0$

\subsection{$Z_8-II-b\ \ \ \theta=(1,3,-4)/8\ \ \ SO(9)\times SO(4)$}
$\Theta=[C(SO(9)),\ C(SO(4))]$

\emph{\tw\ $Z_2$ fixed plane in the $\theta^2$ twisted sector.}

$A_1\approx A_2\approx A_3\approx 2A_4\approx 2A_5\approx 2A_6\approx 0$

\subsection{$Z_8-II-c\ \ \ \theta=(1,3,-4)/8\ \ \ SO(8)^{[2]}\times SO(4)$}
$\Theta=[C(SO(8)^{[2]}),\ C(SO(4))]$

\emph{\tw\ $Z_2$ fixed plane in the $\theta^2$ twisted sector.}

$A_1\approx A_2\approx A_3\approx A_4,\,2A_1\approx 2A_5\approx 2A_6\approx 0$

\subsection{$Z_{12}-I-a\ \ \ \theta=(1,-5,4)/12\ \ \ E_6$}
$\Theta=[C(E_6)]$

\emph{\non\ $Z_3$ fixed plane in the $\theta^3$ twisted sector.}

$A_1\approx A_2\approx A_3\approx A_4\approx A_5,\ 3A_5\approx A_6\approx 0$

\subsection{$Z_{12}-I-b\ \ \ \theta=(1,-5,4)/12\ \ \ F_4\times SU(3)$}
$\Theta=[C(F_4),\ C(SU(3))]$

\emph{\tw\ $Z_3$ fixed plane in the $\theta^3$ twisted sector.}

$A_1\approx A_2\approx A_3\approx A_4\approx 0,\  A_5\approx A_6,\
3A_6\approx 0$

\subsection{$Z_{12}-I-c\ \ \ \theta=(1,-5,4)/12\ \ \ SO(8)^{[3]}\times SU(3)$}
$\Theta=[C(SO(8)^{[3]}),\ C(SU(3))]$

\emph{\tw\ $Z_3$ fixed plane in the $\theta^3$ twisted sector.}

$A_1\approx A_2\approx A_3\approx A_4\approx 0,\  A_5\approx A_6,\
3A_6\approx 0$

\subsection{$Z_{12}-II-a\ \ \ \theta=(1,5,-6)/12\ \ \ F_4\times SO(4)$}
$\Theta=[C(F_4),\ C(SO(4))]$

\emph{\tw\ $Z_2$ fixed plane in the $\theta^2$ twisted sector.}

$A_1\approx A_2\approx A_3\approx A_4\approx 0,\  2A_5\approx 2A_6$

\subsection{$Z_{12}-II-b\ \ \ \theta=(1,5,-6)/12\ \ \ SO(8)^{[3]}\times SO(4)$}
$\Theta=[C(SO(8)^{[3]}),\ C(SO(4))]$

\emph{\tw\ $Z_2$ fixed plane in the $\theta^2$ twisted sector.}

$A_1\approx A_2\approx A_3\approx A_4\approx 0,\  2A_5\approx 2A_6$

\subsection{$Z_2\times Z_2\ \ \ \theta=(1,0,-1)/2\ \ \omega=(0,1,-1)/2\ \ \ 
(SO(4))^3$}
$\Theta=[C(SO(4)),\ I,\  C(SO(4))]\ \ \Omega=[I,\ C(SO(4)),\ C(SO(4))]$

\emph{\tw\ $(I,Z_2),\ (Z_2,I)\ and\ (Z_2,Z_2)$ fixed planes in the 
$\theta,\ \omega$ and $\theta\omega$ twisted sectors respectively.}

$2A_1\approx 2A_2\approx 2A_3\approx 2A_4\approx 2A_5\approx 2A_6\approx 0$

\subsection{$Z_3\times Z_3\ \ \ \theta=(1,0,-1)/3\ \ \omega=(0,1,-1)/3\ \ \ 
(SU(3))^3$}
$\Theta=[C(SU(3)),\ I,\  C(SU(3))]\ \ \Omega=[I,\ C(SU(3)),\ C(SU(3))]$

\emph{\tw\ $(I,Z_3),\ (Z_3,I)\ and\ (Z_3,Z_3)$ fixed planes in the 
$\theta^k,\ \omega^k$ and $(\theta\omega^2)^k$ twisted sectors respectively.}

$A_1\approx A_2,\, A_3\approx A_4,\, A_5\approx A_6,\ 
3A_1\approx 3A_3\approx 3A_5\approx 0$

\subsection{$Z_4\times Z_4\ \ \ \theta=(1,0,-1)/4\ \ \omega=(0,1,-1)/4\ \ \ 
(SO(5))^3$}
$\Theta=[C(SO(5)),\ I,\  C(SO(5))]\ \ \Omega=[I,\ C(SO(5)),\ C(SO(5))]$

\emph{\tw\ $(I,Z_4),\ (Z_4,I)\ and\ (Z_4,Z_4)$ fixed planes in the 
$\theta^k,\ \omega^k$ and $(\theta^2\omega^2)^k$ twisted sectors respectively.}

$A_1\approx A_3\approx A_5\approx 0,\ 
2A_2\approx 2A_4\approx 2A_6\approx 0$

\subsection{$Z_3\times Z_6-a\ \ \ \theta=(1,0,-1)/3\ \ \omega=(0,1,-1)/6\ \ \ 
(SU(3))^3$}
$\Theta=[C(SU(3)),\ I,\  C(SU(3))]\ \ \Omega=[I,\ C(SU(3)^{[2]}),\ 
C(SU(3)^{[2]})]$

\emph{\tw\ $(I,Z_6),\ (Z_3,I)\ and\ (Z_3,Z_6)$ fixed planes in the 
$\theta^k,\ \omega^k$ and $(\theta\omega^4)^k$ twisted sectors respectively.}

$A_1\approx A_2,\ 3A_1\approx 0,\ 
A_3\approx A_4\approx A_5\approx A_6\approx 0$

\subsection{$Z_3\times Z_6-b\ \ \ \theta=(1,0,-1)/3\ \ \omega=(0,1,-1)/6\ \ \ 
SU(3)\times G_2\times SU(3)$}
$\Theta=[C(SU(3)),\ I,\  C(SU(3))]\ \ \Omega=[I,\ C(G_2),\ 
C(SU(3)^{[2]})]$

\emph{\tw\ $(I,Z_6),\ (Z_3,I)\ and\ (Z_3,Z_6)$ fixed planes in the 
$\theta^k,\ \omega^k$ and $(\theta\omega^4)^k$ twisted sectors respectively.}

$A_1\approx A_2,\ 3A_1\approx 0,\ 
A_3\approx A_4\approx A_5\approx A_6\approx 0$

\subsection{$Z_6\times Z_6-a\ \ \ \theta=(1,0,-1)/6\ \ \omega=(0,1,-1)/6\ \ \ 
(G_2)^3$}
$\Theta=[C(G_2),\ I,\  C(G_2)]\ \ \Omega=[I,\ C(G_2),\ C(G_2)]$

\emph{\tw $(I,Z_6),\ (Z_6,I)\ and\ (Z_6,Z_6)$ fixed planes in the 
$\theta^k,\ \omega^k$ and $(\theta\omega^5)^k$ twisted sectors respectively.}

$A_1\approx A_2\approx A_3\approx A_4\approx A_5\approx A_6\approx 0$

\subsection{$Z_6\times Z_6-b\ \ \ \theta=(1,0,-1)/6\ \ \omega=(0,1,-1)/6\ \ \ 
(G_2)^2\times SU(3)$}
$\Theta=[C(G_2),\ I,\  C(SU(3)^{[2]})]\ \ \Omega=[I,\ C(G_2),\ C(SU(3)^{[2]})]$

\emph{\tw\ $(I,Z_6),\ (Z_6,I)\ and\ (Z_6,Z_6)$ fixed planes in the 
$\theta^k,\ \omega^k$ and $(\theta\omega^5)^k$ twisted sectors respectively.}

$A_1\approx A_2\approx A_3\approx A_4\approx A_5\approx A_6\approx 0$

\subsection{$Z_6\times Z_6-c\ \ \ \theta=(1,0,-1)/6\ \ \omega=(0,1,-1)/6\ \ \ 
G_2\times (SU(3))^2$}
$\Theta=[C(G_2),\ I,\  C(SU(3)^{[2]})]\ \ \Omega=[I,\ C(SU(3)^{[2]}),\ 
C(SU(3)^{[2]})]$

\emph{\tw\ $(I,Z_6),\ (Z_6,I)\ and\ (Z_6,Z_6)$ fixed planes in the 
$\theta^k,\ \omega^k$ and $(\theta\omega^5)^k$ twisted sectors respectively.}

$A_1\approx A_2\approx A_3\approx A_4\approx A_5\approx A_6\approx 0$

\subsection{$Z_6\times Z_6-d\ \ \ \theta=(1,0,-1)/6\ \ \omega=(0,1,-1)/6\ \ \ 
(SU(3))^3$}
$\Theta=[C(SU(3)^{[2]}),\ I,\  C(SU(3)^{[2]})]\ \ 
\Omega=[I,\ C(SU(3)^{[2]}),\ C(SU(3)^{[2]})]$

\emph{\tw $(I,Z_6),\ (Z_6,I)\ and\ (Z_6,Z_6)$ fixed planes in the 
$\theta^k,\ \omega^k$ and $(\theta\omega^5)^k$ twisted sectors respectively.}

$A_1\approx A_2\approx A_3\approx A_4\approx A_5\approx A_6\approx 0$

\section{Modular Symmetries of New Orbifolds}

We now present the calculation of the modular symmetries and threshold 
corrections for $Z_6-II-f$ without Wilson lines. The derivation uses
the methods applied earlier \cite{may1,bai2,bai3} to Coxeter $Z_N$ orbifolds. 
The matrix $Q$ representing the action of the point group is $\Theta^T$ where
\begin{eqnarray}
\Theta =(C(SU(4)^{[2]}),C(SU(3)),C(SU(2))) \end{eqnarray}
as in appendix B. There is a \non\ $Z_2$ fixed plane in the $\theta^2$ twisted
sector. Solving
\begin{eqnarray} Q^2w=w \end{eqnarray}
and
\begin{eqnarray} (Q^*)^2p=p \end{eqnarray}
where
\begin{eqnarray} Q^*\equiv(Q^T)^{-1} \end{eqnarray}
gives the solutions $W_{soln}$ and $p_{soln}$ for the winding numbers and 
momenta in the $\theta^2$ twisted sector
\begin{eqnarray} w_{soln}=\left(\matrix{n_1\cr 2n_1\cr -n_1\cr 0\cr 0\cr n_6\cr}
\right) \end{eqnarray}
and
\begin{eqnarray} p_{soln}=\left(\matrix{0\cr m_2\cr -m_2\cr 0\cr 0\cr m_6\cr}
\right) \end{eqnarray} by
The world sheet momentum $P$ is given by 
\begin{eqnarray} P=p^T_{soln}w_{soln}=m^T_{\perp}K^Tn_{\perp} \end{eqnarray}
where
\begin{eqnarray} m_{\perp}=\left(\matrix{m_2\cr m_6\cr}\right) \,,\ n_{\perp}=
\left(\matrix{n_1\cr n_6\cr}\right) \end{eqnarray}
Thus,
\begin{eqnarray} K=\left( \matrix{3&0\cr 0&1\cr}\right) \end{eqnarray}
This matrix $K$ has the same form as for the $\theta^2$ twisted sector of the
$Z_6$--II--c orbifold \cite{bai2,bai3}
despite the very different form of the $w_{soln}$
and $p_{soln}$. Thus the modular symmetry group associated with the $Z_2$ 
fixed planes is $\Gamma^0(3)_{T}\times \Gamma_0(3)_{U}$ in the notation of 
earlier work \cite{may1,bai2,bai3} for congruence subgroups of PSL(2,Z), 
where the first factor refers to the $T$ modulus and the second factor to the
$U$ modulus.

In addition, because we are dealing with the same orbifold appart from the 
lattice, the fundamental sectors and the construction of the renormalisation 
group co--efficient $b_a^{N=2}$ from the co--efficients $b_a^{(I,\theta^2)},
b_a^{(\theta^4,\theta^2)},b_a^{(\theta^4,\theta^4)}$ and $b_a^{(\theta^2,I)}$
for the various twisted sectors is the same as for $Z_6$--II--c. With $K$
as above, the matrix $A$ involved in the construction of the partition
function is 
\begin{eqnarray} A=\left(\matrix{n_1&l_1/3\cr n_2&l_2 \cr}\right) \end{eqnarray}
just as for $Z_6$--II--c, and we obtain for the contribution of these sectors 
to the threshold correction
\begin{eqnarray} \Delta_a=-2b_a^{(I,\theta^2)}ln(kT_2\vert \eta(T)\vert^4 
U_2\vert \eta(U)\vert^4)-2b_a^{(I,\theta^2)}ln(kT_2\vert \eta(T/3)\vert^4 
U_2\vert \eta(3U)\vert^4) \end{eqnarray}

\section{Modular Subgroups for \tw\ Orbifolds}

We now list all the obtainable modular symmetries
for $Z_2$,$Z,3$ and $Z_4$ planes, where the plane
lies completely in the $T^2$ sublattice, for all choices of Wilson lines
consistent with modular and point group invariance.
For each modular group we also provide an example Wilson line which yields this
symmetry.

\subsection{$Z_2$ Plane}

For the $T$ modulus there are 7 possible modular groups:

\textbf{1.}\ PSL(2,Z)

\ \ $A=\mathbf{0}$

\textbf{2.}\ $a,d=1 \bmod 4$, $c=0 \bmod 4$ and $a,d=3 \bmod 4$, $c=0 \bmod 4$  

\ \ $A={1\over 2}\left(\matrix{0&0&0&0&0&0&0&0&0&0&0&0&0&0&0&0\cr 
0&0&0&0&0&0&0&0&1&1&0&0&0&0&0&0\cr}\right)$

\textbf{3.}\ $a,d=1\bmod 2$, $c=0\bmod 2$

\ \ $A={1\over 2}\left(\matrix{0&0&0&0&0&0&0&0&0&0&0&0&0&0&0&0\cr 
0&0&0&0&0&0&0&0&2&0&0&0&0&0&0&0\cr}\right)$

\textbf{4.}\ $a,d=1 \bmod 4$, $c=0 \bmod 16$ and $a,d=3 \bmod 4$, $c=0 \bmod 16$

\ \ $A={1\over 2}\left(\matrix{1&1&0&0&0&0&0&0&0&0&0&0&0&0&0&0\cr 
0&0&0&0&0&0&0&0&1&1&0&0&0&0&0&0\cr}\right)$

\textbf{5.}\ $a,d=1 \bmod 4$, $c=0 \bmod 8$ and $a,d=3 \bmod 4$, $c=0 \bmod 8$

\ \ $A={1\over 2}\left(\matrix{1&1&0&0&0&0&0&0&0&0&0&0&0&0&0&0\cr 
0&0&0&0&0&0&0&0&2&0&0&0&0&0&0&0\cr}\right)$

\textbf{6.}\ $a,d=1\bmod 2$, $c=0\bmod 4$

\ \ $A={1\over 2}\left(\matrix{1&1&0&0&0&0&0&0&1&1&0&0&0&0&0&0\cr 
0&0&0&0&0&0&0&0&0&0&2&0&0&0&0&0\cr}\right)$

\textbf{7.}\ $a,d=1\bmod 4$, $c=0\bmod 16$ and $a=1\bmod 4$, $d=3\bmod 4$,
$c=8\bmod 16$
and $a,d=3\bmod 4$, $c=0\bmod 16$ and $a=3\bmod 4$, $d=1\bmod 4$, $c=8\bmod 16$

\ \ $A={1\over 2}\left(\matrix{1&1&0&0&0&0&0&0&1&1&0&0&0&0&0&0\cr 
1&0&1&0&0&0&0&0&1&0&1&0&0&0&0&0\cr}\right)$

For the $U$ modulus there are 10 modular groups:

\textbf{1.}\ PSL(2,Z)

\ \ $A=\mathbf{0}$

\textbf{2.}\ $a,d=1 \bmod 4$, $c=0 \bmod 4$ and $a,d=3 \bmod 4$, $c=0 \bmod 4$  

\ \ $A={1\over 2}\left(\matrix{0&0&0&0&0&0&0&0&0&0&0&0&0&0&0&0\cr 
0&0&0&0&0&0&0&0&1&1&0&0&0&0&0&0\cr}\right)$

\textbf{3.}\ $a,d=1\bmod 2$, $c=0\bmod 2$

\ \ $A={1\over 2}\left(\matrix{0&0&0&0&0&0&0&0&0&0&0&0&0&0&0&0\cr 
0&0&0&0&0&0&0&0&2&0&0&0&0&0&0&0\cr}\right)$

\textbf{4.}\ $a,d=1 \bmod 4$, $b=0 \bmod 4$ and $a,d=3 \bmod 4$, $b=0 \bmod 4$  

\ \ $A={1\over 2}\left(\matrix{1&1&0&0&0&0&0&0&0&0&0&0&0&0&0&0\cr 
0&0&0&0&0&0&0&0&0&0&0&0&0&0&0&0\cr}\right)$

\textbf{5.}\ $a,d=1\bmod 4$, $b,c=0\bmod 4$ and $a,d=1\bmod 4$, $b,c=2\bmod 4$ 
and $a,d=3\bmod 4$, $b,c=0\bmod 4$ and $a,d=3\bmod 4$, $b,c=2\bmod 4$

\ \ $A={1\over 2}\left(\matrix{1&1&0&0&0&0&0&0&0&0&0&0&0&0&0&0\cr 
0&0&0&0&0&0&0&0&1&1&0&0&0&0&0&0\cr}\right)$

\textbf{6.}\ $a,d=1\bmod 4$, $b=0\bmod 4$, $c=0\bmod 2$ and $a,d=3\bmod 4$,
$b=0\bmod 4$, $c=0\bmod 2$

\ \ $A={1\over 2}\left(\matrix{1&1&0&0&0&0&0&0&0&0&0&0&0&0&0&0\cr 
0&0&0&0&0&0&0&0&2&0&0&0&0&0&0&0\cr}\right)$

\textbf{7.}\ $a,d=1\bmod 2$, $b=0\bmod 2$

\ \ $A={1\over 2}\left(\matrix{1&1&0&0&0&0&0&0&1&1&0&0&0&0&0&0\cr 
0&0&0&0&0&0&0&0&0&0&0&0&0&0&0&0\cr}\right)$

\textbf{8.}\ $a,d=1\bmod 4$, $b=0\bmod 2$, $c=0\bmod 4$ and $a,d=3\bmod 4$,
$b=0\bmod 2$, $c=0\bmod 4$

\ \ $A={1\over 4}\left(\matrix{2&2&0&0&0&0&0&0&2&2&0&0&0&0&0&0\cr 
0&0&0&0&0&0&0&0&1&-1&-1&-1&-1&1&1&1\cr}\right)$

\textbf{9.}\ $a,d=1\bmod 2$, $b,c=0\bmod 2$

\ \ $A={1\over 2}\left(\matrix{1&1&0&0&0&0&0&0&1&1&0&0&0&0&0&0\cr 
0&0&0&0&0&0&0&0&2&0&0&0&0&0&0&0\cr}\right)$

\textbf{10.}\ $a,d=1\bmod 2$, $b,c=0\bmod 2$ and $a,d=0\bmod 2$, $b,c=1\bmod 2$

\ \ $A={1\over 2}\left(\matrix{2&0&0&0&0&0&0&0&0&0&0&0&0&0&0&0\cr 
2&0&0&0&0&0&0&0&0&0&0&0&0&0&0&0\cr}\right)$

\subsection{$Z_3$ Plane}

The $U$ modulus is fixed for a $Z_3$ plane.
For the $T$ modulus there are 2 modular groups:

\textbf{1.}\ PSL(2,Z)

\ \ $A=\mathbf{0}$

\textbf{2.}\ $a,d=1\bmod 3$, $c=0\bmod 3$

\ \ $A={1\over 3}\left(\matrix{2&1&1&0&0&0&0&0&0&0&0&0&0&0&0&0\cr 
2&1&1&0&0&0&0&0&0&0&0&0&0&0&0&0\cr}\right)$

\subsection{$Z_4$ Plane}

The $U$ modulus is fixed for a $Z_4$ plane.
For the $T$ modulus there are 2 modular groups:

\textbf{1.}\ PSL(2,Z)

\ \ $A=\mathbf{0}$

\textbf{2.}\ $\Gamma_0(2)$

\ \ $A={1\over 2}\left(\matrix{0&0&0&0&0&0&0&0&0&0&0&0&0&0&0&0\cr 
2&0&0&0&0&0&0&0&0&0&0&0&0&0&0&0\cr}\right)$

\section{Modular Subgroups for \non\ Orbifolds}

We now list all the obtainable modular symmetries
of fixed planes, where the orbifold cannot be decomposed as \tw\, 
for all choices of Wilson lines
consistent with modular and point group invariance.
For each modular group we also provide an example Wilson line which yields this
symmetry.

\subsection{$Z_4$--a $(SU(4)^2)$}
There is a \non\ $Z_2$ fixed plane in the $\theta^2$ twisted sector.

$R=\left(\matrix{2&0\cr 0&2\cr}\right)\ \ K=\left(\matrix{2&0\cr 0&2\cr}\right)$

Wilson lines as stated here correspond to thes six component line as:
$A=\left(\matrix{A_1\cr A_4\cr}\right)$

For the $T$ modulus there are 6 modular groups:

\textbf{1.}\ $\Gamma^0(2)$

\ \ $A=\mathbf{0}$

\textbf{2.}\ $a=1\bmod 4$, $b,c=0\bmod 2$, $d=1\bmod 2$

\ \ $A={1\over 4}\left(\matrix{0&0&0&0&0&0&0&0&0&0&0&0&0&0&0&0\cr 
2&1&1&1&1&0&0&0&0&0&0&0&0&0&0&0\cr}\right)$

\textbf{3.}\ $a=1\bmod 4$, $b=0\bmod 2$, $c=0\bmod 4$, $d=1\bmod 2$ and
$a=3\bmod 4$, $b=0\bmod 2$, $c=2\bmod 4$, $d=1\bmod 2$

\ \ $A={1\over 4}\left(\matrix{2&0&0&0&0&0&0&0&2&0&0&0&0&0&0&0\cr 
0&2&2&2&0&0&0&0&0&2&2&2&0&0&0&0\cr}\right)$

\textbf{4.}\ $a=1\bmod 4$, $b=0\bmod 2$, $c=0\bmod 4$, $d=1\bmod 2$

\ \ $A={1\over 4}\left(\matrix{2&1&1&0&0&0&0&0&1&1&0&0&0&0&0&0\cr 
0&0&0&0&0&0&0&0&1&-1&1&1&1&1&1&1\cr}\right)$

\textbf{5.}\ $a,d=1\bmod 2$, $b,c=0\bmod 2$

\ \ $A={1\over 4}\left(\matrix{2&2&0&0&0&0&0&0&0&0&0&0&0&0&0&0\cr 
0&0&0&0&0&0&0&0&2&2&0&0&0&0&0&0\cr}\right)$

\textbf{6.}\ $a,d=0\bmod 2$, $b,c=1\bmod 2$ and $a,d=1\bmod 2$, $b,c=0\bmod 2$

\ \ $A={1\over 4}\left(\matrix{2&2&0&0&0&0&0&0&2&2&0&0&0&0&0&0\cr 
-1&-1&3&-1&-1&1&1&-1&-1&-1&3&-1&-1&1&1&-1\cr}\right)$

For the $U$ modulus there are 9 modular groups:

\textbf{1.}\ PSL(2,Z)

\ \ $A=\mathbf{0}$

\textbf{2.}\ $a,d=1\bmod 2$, $c=0\bmod 2$

\ \ $A={1\over 4}\left(\matrix{0&0&0&0&0&0&0&0&0&0&0&0&0&0&0&0\cr 
2&2&0&0&0&0&0&0&0&0&0&0&0&0&0&0\cr}\right)$

\textbf{3.}\ $a=1\bmod 4$, $c=0\bmod 2$, $d=1\bmod 2$

\ \ $A={1\over 4}\left(\matrix{0&0&0&0&0&0&0&0&0&0&0&0&0&0&0&0\cr 
2&1&1&1&1&0&0&0&0&0&0&0&0&0&0&0\cr}\right)$

\textbf{4.}\ $a=1\bmod 2$, $b=0\bmod 4$, $d=1\bmod 4$

\ \ $A={1\over 4}\left(\matrix{2&0&0&0&0&0&0&0&2&0&0&0&0&0&0&0\cr 
0&0&0&0&0&0&0&0&0&0&0&0&0&0&0&0\cr}\right)$

\textbf{5.}\ $a,d=1\bmod 4$, $b,c=0\bmod 2$ and
$a=3\bmod 4$, $b=2\bmod 4$, $c=0\bmod 4$, $d=1\bmod 4$ and
$a=1\bmod 4$, $b=0\bmod 4$, $c=2\bmod 4$, $d=3\bmod 4$ and
$a=3\bmod 4$, $b=2\bmod 4$, $c=2\bmod 4$, $d=3\bmod 4$

\ \ $A={1\over 4}\left(\matrix{2&0&0&0&0&0&0&0&2&0&0&0&0&0&0&0\cr 
0&2&2&2&0&0&0&0&0&2&2&2&0&0&0&0\cr}\right)$

\textbf{6.}\ $a=1\bmod 2$, $b=0\bmod 4$, $c=0\bmod 2$, $d=1\bmod 4$

\ \ $A={1\over 4}\left(\matrix{2&1&1&0&0&0&0&0&1&1&0&0&0&0&0&0\cr 
0&0&0&0&0&0&0&0&1&-1&1&1&1&1&1&1\cr}\right)$

\textbf{7.}\ $a,d=1\bmod 4$, $b,d=0\bmod 4$

\ \ $A={1\over 8}\left(\matrix{4&2&2&0&0&0&0&0&2&2&0&0&0&0&0&0\cr 
-1&1&1&-1&-1&1&1&-1&0&0&2&-2&-2&2&2&-2\cr}\right)$

\textbf{8.}\ $a,d=1\bmod 2$, $b=0\bmod 2$

\ \ $A={1\over 4}\left(\matrix{2&2&0&0&0&0&0&0&0&0&0&0&0&0&0&0\cr 
0&0&0&0&0&0&0&0&0&0&0&0&0&0&0&0\cr}\right)$

\textbf{9.}\ $a,d=1\bmod 2$, $b,c=0\bmod 2$

\ $A={1\over 4}\left(\matrix{2&2&0&0&0&0&0&0&0&0&0&0&0&0&0&0\cr 
0&0&0&0&0&0&0&0&2&2&0&0&0&0&0&0\cr}\right)$

\subsection{$Z_4$--b $SU(4)\times SO(5)\times SU(2)$}
There is a \non\ $Z_2$ fixed plane in the $\theta^2$ twisted sector.

$R=\left(\matrix{2&0\cr 0&1\cr}\right)\ \ K=\left(\matrix{2&0\cr 0&1\cr}\right)$

Wilson lines as stated here correspond to thes six component line as:
$A=\left(\matrix{A_1\cr A_6\cr}\right)$

For the $T$ modulus there are 9 modular groups:

\textbf{1.}\ $\Gamma^0(2)$

\ \ $A=\mathbf{0}$

\textbf{2.}\ $a,d=1\bmod 4$, $b=0\bmod 2$, $c=0\bmod 4$ and 
$a,d=3\bmod 4$, $b=0\bmod 2$, $c=0\bmod 4$

\ \ $A={1\over 2}\left(\matrix{0&0&0&0&0&0&0&0&0&0&0&0&0&0&0&0\cr 
1&1&0&0&0&0&0&0&0&0&0&0&0&0&0&0\cr}\right)$

\textbf{3.}\ $a=1\bmod 2$, $b,c=0\bmod 2$

\ \ $A={1\over 4}\left(\matrix{0&0&0&0&0&0&0&0&0&0&0&0&0&0&0&0\cr 
2&0&0&0&0&0&0&0&0&0&0&0&0&0&0&0\cr}\right)$

\textbf{4.}\ $a,d=1\bmod 4$, $b,c=0\bmod 2$

\ \ $A={1\over 4}\left(\matrix{2&0&0&0&0&0&0&0&2&0&0&0&0&0&0&0\cr 
0&0&0&0&0&0&0&0&0&0&0&0&0&0&0&0\cr}\right)$

\textbf{5.}\ $a,d=1\bmod 4$, $b=0\bmod 2$, $c=0\bmod 8$

\ $A={1\over 4}\left(\matrix{2&0&0&0&0&0&0&0&2&0&0&0&0&0&0&0\cr 
0&1&1&0&0&0&0&0&0&0&0&0&0&0&0&0\cr}\right)$

\textbf{6.}\ $a=1\bmod 4$, $b=0\bmod 2$, $c=0\bmod 4$, $d=1\bmod 2$

\ \ $A={1\over 4}\left(\matrix{2&0&0&0&0&0&0&0&2&0&0&0&0&0&0&0\cr 
0&2&0&0&0&0&0&0&0&0&0&0&0&0&0&0\cr}\right)$

\textbf{7.}\ \ $a=1\bmod 4$, $b=0\bmod 2$, $c=0\bmod 4$, $d=1\bmod 2$ and
$a=3\bmod 4$, $b=0\bmod 2$, $c=2\bmod 4$, $d=1\bmod 2$

\ $A={1\over 4}\left(\matrix{2&0&0&0&0&0&0&0&2&0&0&0&0&0&0&0\cr 
0&2&0&0&0&0&0&0&0&2&0&0&0&0&0&0\cr}\right)$

\textbf{8.}\ $a,d=1\bmod 4$, $b=0\bmod 2$, $c=0\bmod 8$ and
$a,d=3\bmod 4$, $b=0\bmod 2$, $c=0\bmod 8$ and
$a,d=1\bmod 4$, $b=1\bmod 2$, $c=4\bmod 8$ and
$a,d=3\bmod 4$, $b=1\bmod 2$, $c=4\bmod 8$

\ $A={1\over 4}\left(\matrix{2&2&0&0&0&0&0&0&0&0&0&0&0&0&0&0\cr 
0&0&0&0&0&0&0&0&1&1&0&0&0&0&0&0\cr}\right)$

\textbf{9.}\ $a,d=1\bmod 2$, $b=0\bmod 2$, $c=0\bmod 4$

\ \ $A={1\over 4}\left(\matrix{2&2&0&0&0&0&0&0&0&0&0&0&0&0&0&0\cr 
0&0&0&0&0&0&0&0&2&0&0&0&0&0&0&0\cr}\right)$

\subsection{$Z_6$--II--a $SU(6)\times SU(2)$}
There is a \non\ $Z_2$ fixed plane in the $\theta^2$ twisted sector and
a \non\ $Z_3$ fixed plane in the $\theta^3$ twisted sector.

$R=\left(\matrix{3&0\cr 0&1\cr}\right)\ \ K=\left(\matrix{3&0\cr 0&1\cr}\right)$

Wilson lines as stated here correspond to thes six component line as:
$A=\left(\matrix{A_1\cr A_6\cr}\right)$

\emph{The $\theta^2$ sector:}

For the $T$ modulus there are 6 modular groups:

\textbf{1.}\ $\Gamma^0(3)$

\ \ $A=\mathbf{0}$

\textbf{2.}\ $a,d=1\bmod 4$, $b=0\bmod 3$, $c=0\bmod 4$ and 
$a,d=3\bmod 4$, $b=0\bmod 3$, $c=0\bmod 4$

\ \ $A={1\over 2}\left(\matrix{0&0&0&0&0&0&0&0&0&0&0&0&0&0&0&0\cr 
1&1&0&0&0&0&0&0&0&0&0&0&0&0&0&0\cr}\right)$

\textbf{3.}\ $a,d=1\bmod 2$, $b=0\bmod 3$, $c=0\bmod 2$

\ \ $A={1\over 2}\left(\matrix{0&0&0&0&0&0&0&0&0&0&0&0&0&0&0&0\cr 
2&0&0&0&0&0&0&0&0&0&0&0&0&0&0&0\cr}\right)$

\textbf{4.}\ $a=1\bmod 6$, $b=0\bmod 3$, $c=0\bmod 2$, $d=1\bmod 2$

\ \ $A={1\over 6}\left(\matrix{2&1&1&0&0&0&0&0&2&1&1&0&0&0&0&0\cr 
0&0&0&0&0&0&0&0&0&0&0&0&0&0&0&0\cr}\right)$

\textbf{5.}\ $a=1\bmod 12$, $b=0\bmod 3$, $c=0\bmod 8$, $d=1\bmod 4$ and
$a=7\bmod 12$, $b=0\bmod 3$, $c=0\bmod 8$, $d=3\bmod 4$

\ \ $A={1\over 12}\left(\matrix{4&2&2&0&0&0&0&0&4&2&2&0&0&0&0&0\cr 
0&0&0&0&0&0&0&0&-3&3&3&-3&-3&3&3&-3\cr}\right)$

\textbf{6.}\ $a=1\bmod 6$, $b=0\bmod 3$, $c=0\bmod 8$, $d=1\bmod 2$

\ \ $A={1\over 12}\left(\matrix{4&2&2&0&0&0&0&0&4&2&2&0&0&0&0&0\cr 
0&0&0&0&0&0&0&0&-3&9&-3&-3&-3&3&3&-3\cr}\right)$

For the $U$ modulus there are 7 modular groups:

\textbf{1.}\ $\Gamma_0(3)$

\ \ $A=\mathbf{0}$

\textbf{2.}\ $a,d=1\bmod 4$, $c=0\bmod 12$ and 
$a,d=3\bmod 4$, $c=0\bmod 12$

\ \ $A={1\over 2}\left(\matrix{0&0&0&0&0&0&0&0&0&0&0&0&0&0&0&0\cr 
1&1&0&0&0&0&0&0&0&0&0&0&0&0&0&0\cr}\right)$

\textbf{3.}\ $a,d=1\bmod 2$, $c=0\bmod 6$

\ \ $A={1\over 2}\left(\matrix{0&0&0&0&0&0&0&0&0&0&0&0&0&0&0&0\cr 
2&0&0&0&0&0&0&0&0&0&0&0&0&0&0&0\cr}\right)$

\textbf{4.}\ $a=1\bmod 6$, $b=0\bmod 2$, $c=0\bmod 3$, $d=1\bmod 12$

\ \ $A={1\over 6}\left(\matrix{2&1&1&0&0&0&0&0&2&1&1&0&0&0&0&0\cr 
0&0&0&0&0&0&0&0&0&0&0&0&0&0&0&0\cr}\right)$

\textbf{5.}\ $a=1\bmod 4$, $b=0\bmod 2$, $c=0\bmod 12$, $d=1\bmod 12$ and
$a=3\bmod 4$, $b=0\bmod 2$, $c=0\bmod 12$, $d=7\bmod 12$

\ \ $A={1\over 12}\left(\matrix{4&2&2&0&0&0&0&0&4&2&2&0&0&0&0&0\cr 
0&0&0&0&0&0&0&0&-3&3&3&-3&-3&3&3&-3\cr}\right)$

\textbf{6.}\ $a=1\bmod 2$, $b=0\bmod 2$, $c=0\bmod 6$, $d=1\bmod 6$

\ \ $A={1\over 12}\left(\matrix{4&2&2&0&0&0&0&0&4&2&2&0&0&0&0&0\cr 
0&0&0&0&0&0&0&0&-3&9&-3&-3&-3&3&3&-3\cr}\right)$

\textbf{7.}\ $a=1\bmod 2$, $b=0\bmod 2$, $c=0\bmod 6$, $d=1\bmod 6$ and
$a=0\bmod 2$, $b=1\bmod 2$, $c=3\bmod 6$, $d=4\bmod 6$

\ \ $A={1\over 6}\left(\matrix{2&2&2&0&0&0&0&0&0&0&0&0&0&0&0&0\cr 
0&0&0&6&0&0&0&0&0&0&0&0&0&0&0&0\cr}\right)$

\emph{The $\theta^3$ sector:}

$R=\left(\matrix{2&0\cr 0&2\cr}\right)\ \ K=\left(\matrix{2&0\cr 0&2\cr}\right)$

Wilson lines as stated here correspond to thes six component line as:
$A=\left(\matrix{A_1\cr A_4\cr}\right)$

For the $T$ modulus there are 3 modular groups:

\textbf{1.}\ $\Gamma^0(2)$

\ \ $A=\mathbf{0}$

\textbf{2.}\ $a=1\bmod 6$, $b=0\bmod 2$, $c=0\bmod 3$, $d=1\bmod 3$

\ \ $A={1\over 6}\left(\matrix{2&1&1&0&0&0&0&0&2&1&1&0&0&0&0&0\cr 
2&1&1&0&0&0&0&0&2&1&1&0&0&0&0&0\cr}\right)$

\textbf{3.}\ $a,d=1\bmod 3$, $b=0\bmod 2$, $c=0\bmod 3$

\ \ $A={1\over 6}\left(\matrix{4&0&0&0&0&0&0&0&2&2&0&0&0&0&0&0\cr 
4&0&0&0&0&0&0&0&2&2&0&0&0&0&0&0\cr}\right)$

\subsection{$Z_6$--II--b $SU(3)\times SO(8)$}
There is a \non\ $Z_2$ fixed plane in the $\theta^2$ twisted sector.

$R=\left(\matrix{0&-1\cr -1&0\cr}\right)\ \ 
K=\left(\matrix{2&1\cr 1&2\cr}\right)$

Wilson lines as stated here correspond to thes six component line as:
$A=\left(\matrix{A_5\cr A_6\cr}\right)$

For the $T$ modulus there are 4 modular groups:

\textbf{1.}\ $\Gamma^0(3)$

\ \ $A=\mathbf{0}$

\textbf{2.}\ $a,d=1\bmod 2$, $b=0\bmod 3$, $c=0\bmod 2$

\ \ $A={1\over 2}\left(\matrix{0&0&0&0&0&0&0&0&0&0&0&0&0&0&0&0\cr 
2&0&0&0&0&0&0&0&0&0&0&0&0&0&0&0\cr}\right)$

\textbf{3.}\ $a,d=1\bmod 4$, $b=0\bmod 3$, $c=0\bmod 8$ and 
\ $a,d=3\bmod 4$, $b=0\bmod 3$, $c=0\bmod 8$

\ \ $A={1\over 2}\left(\matrix{1&1&0&0&0&0&0&0&0&0&0&0&0&0&0&0\cr 
2&0&0&0&0&0&0&0&1&1&0&0&0&0&0&0\cr}\right)$

\textbf{4.}\ $a,d=1\bmod 2$, $b=0\bmod 3$, $c=0\bmod 4$

\ \ $A={1\over 2}\left(\matrix{1&1&0&0&0&0&0&0&0&0&0&0&0&0&0&0\cr 
0&0&0&0&0&0&0&0&0&0&0&0&0&0&0&0\cr}\right)$

For the $U$ modulus there are 4 modular groups:

\textbf{1.}\ ${1\over 3}a+{2\over 3}b-{2\over 3}c-{1\over 3}d\in Z$,
${2\over 3}a+{1\over 3}b-{1\over 3}c-{2\over 3}d \in Z$

\ \ $A=\mathbf{0}$

\textbf{2.}\ ${1\over 3}a+{2\over 3}b-{2\over 3}c-{1\over 3}d\in Z$,
${2\over 3}a+{1\over 3}b-{1\over 3}c-{2\over 3}d \in Z$, $a=1\bmod 2$,
$b=0\bmod 2$, $-{1\over 3}a-{1\over 6}b+{2\over 3}c+{1\over 3}d \in Z$,
${1\over 6}a+{1\over 3}b-{1\over 3}c-{1\over 6}d \in Z$,
$-{1\over 2}+{2\over 3}a+{1\over 3}b-{1\over 3}c-{1\over 6}d \in Z$

\ \ $A={1\over 2}\left(\matrix{0&0&0&0&0&0&0&0&0&0&0&0&0&0&0&0\cr 
2&0&0&0&0&0&0&0&0&0&0&0&0&0&0&0\cr}\right)$

\textbf{3.}\ ${1\over 3}a+{2\over 3}b-{2\over 3}c-{1\over 3}d\in Z$,
${2\over 3}a+{1\over 3}b-{1\over 3}c-{2\over 3}d \in Z$, $a=1\bmod 2$,
$b=0\bmod 2$, $d=1\bmod 2$, $c=0\bmod 2$,
$-{1\over 2}-{1\over 3}a-{5\over 12}b+{11\over 12}c+{5\over 6}d \in Z$,
$-{1\over 2}-{1\over 4}a-{1\over 2}b+{1\over 2}c+{3\over 4}d \in Z$,
$-{1\over 2}-{11\over 12}a+{5\over 6}b-{5\over 6}c-{5\over 12}d \in Z$,
$-{1\over 2}+{1\over 2}a-{1\over 4}b-{1\over 4}c+\in Z$,
${1\over 2}+{1\over 6}a+{1\over 3}b-{1\over 3}c-{2\over 3}d \in Z$,
${1\over 3}a-{2\over 3}b+{1\over 6}c+{1\over 3}d \in Z$,
$-{1\over 3}a-{1\over 6}b+{2\over 3}c+{1\over 3}d \in Z$,
$-{1\over 2}+{2\over 3}a+{1\over 3}b-{1\over 3}c-{1\over 6}d \in Z$

\ \ $A={1\over 2}\left(\matrix{1&1&0&0&0&0&0&0&0&0&0&0&0&0&0&0\cr 
2&0&0&0&0&0&0&0&1&1&0&0&0&0&0&0\cr}\right)$

\textbf{4.}\ ${1\over 3}a+{2\over 3}b-{2\over 3}c-{1\over 3}d\in Z$,
${2\over 3}a+{1\over 3}b-{1\over 3}c-{2\over 3}d \in Z$, $c=0\bmod 2$,
$d=1\bmod 2$, ${1\over 3}a+{2\over 3}b-{1\over 6}c-{1\over 3}d\in Z$,
${1\over 2}+{1\over 6}a+{1\over 3}b-{1\over 3}c-{2\over 3}d\in Z$,

\ \ $A={1\over 2}\left(\matrix{1&1&0&0&0&0&0&0&0&0&0&0&0&0&0&0\cr 
0&0&0&0&0&0&0&0&0&0&0&0&0&0&0&0\cr}\right)$

\subsection{$Z_6$--II--c $SU(3)\times SO(7)\times SU(2)$}
There is a \non\ $Z_2$ fixed plane in the $\theta^2$ twisted sector.

$R=\left(\matrix{1&0\cr 0&1\cr}\right)\ \ K=\left(\matrix{3&0\cr 0&1\cr}\right)$

Wilson lines as stated here correspond to thes six component line as:
$A=\left(\matrix{A_5\cr A_6\cr}\right)$

For the $T$ modulus there are 5 modular groups:

\textbf{1.}\ $\Gamma^0(3)$

\ \ $A=\mathbf{0}$

\textbf{2.}\ $a=1\bmod 4$, $b=0\bmod 3$, $c=0\bmod 4$, $d=1\bmod 4$ and
$a=3\bmod 4$, $b=0\bmod 3$, $c=0\bmod 4$, $d=3\bmod 4$

\ \ $A={1\over 2}\left(\matrix{0&0&0&0&0&0&0&0&0&0&0&0&0&0&0&0\cr 
1&1&0&0&0&0&0&0&0&0&0&0&0&0&0&0\cr}\right)$

\textbf{3.}\ $a,d=1\bmod 2$, $b=0\bmod 3$, $c=0\bmod 2$

\ \ $A={1\over 2}\left(\matrix{0&0&0&0&0&0&0&0&0&0&0&0&0&0&0&0\cr 
2&0&0&0&0&0&0&0&0&0&0&0&0&0&0&0\cr}\right)$

\textbf{4.}\ $a,d=1\bmod 4$, $b=0\bmod 3$, $c=0\bmod 8$ and
$a,d=3\bmod 4$, $b=0\bmod 3$, $c=0\bmod 8$ and

\ \ $A={1\over 4}\left(\matrix{2&2&0&0&0&0&0&0&2&2&0&0&0&0&0&0\cr 
0&0&0&0&0&0&0&0&1&-1&1&-1&-1&1&1&-1\cr}\right)$

\textbf{5.}\ $a,d=1\bmod 2$, $b=0\bmod 3$, $c=0\bmod 4$

\ \ $A={1\over 2}\left(\matrix{1&1&0&0&0&0&0&0&1&1&0&0&0&0&0&0\cr 
0&0&0&0&0&0&0&0&0&0&2&0&0&0&0&0\cr}\right)$

For the $U$ modulus there are 7 modular groups:

\textbf{1.}\ $\Gamma_0(3)$

\ \ $A=\mathbf{0}$

\textbf{2.}\ $a,d=1\bmod 4$, $c=0\bmod 12$ and 
$a,d=3\bmod 4$, $c=0\bmod 12$

\ \ $A={1\over 2}\left(\matrix{0&0&0&0&0&0&0&0&0&0&0&0&0&0&0&0\cr 
1&1&0&0&0&0&0&0&0&0&0&0&0&0&0&0\cr}\right)$

\textbf{3.}\ $a,d=1\bmod 2$, $c=0\bmod 6$

\ \ $A={1\over 2}\left(\matrix{0&0&0&0&0&0&0&0&0&0&0&0&0&0&0&0\cr 
2&0&0&0&0&0&0&0&0&0&0&0&0&0&0&0\cr}\right)$

\textbf{4.}\ $a=1\bmod 2$, $b=0\bmod 2$, $c=0\bmod 3$, $d=1\bmod 2$

\ \ $A={1\over 2}\left(\matrix{1&1&0&0&0&0&0&0&1&1&0&0&0&0&0&0\cr 
0&0&0&0&0&0&0&0&0&0&0&0&0&0&0&0\cr}\right)$

\textbf{5.}\ $a,d=1\bmod 4$, $b=0\bmod 2$, $c=0\bmod 12$ and
$a,d=3\bmod 4$, $b=0\bmod 2$, $c=0\bmod 12$

\ \ $A={1\over 4}\left(\matrix{2&2&0&0&0&0&0&0&2&2&0&0&0&0&0&0\cr 
0&0&0&0&0&0&0&0&1&-1&1&-1&-1&1&1&-1\cr}\right)$

\textbf{6.}\ $a,d=1\bmod 2$, $b=0\bmod 2$, $c=0\bmod 6$

\ \ $A={1\over 2}\left(\matrix{1&1&0&0&0&0&0&0&1&1&0&0&0&0&0&0\cr 
0&0&0&0&0&0&0&0&0&0&2&0&0&0&0&0\cr}\right)$

\textbf{7.}\ $a,d=0\bmod 2$, $b=1\bmod 2$, $c=3\bmod 6$ and
$a,d=1\bmod 2$, $b=0\bmod 2$, $c=0\bmod 6$

\ \ $A={1\over 2}\left(\matrix{2&0&0&0&0&0&0&0&0&0&0&0&0&0&0&0\cr 
0&2&0&0&0&0&0&0&0&0&0&0&0&0&0&0\cr}\right)$

\subsection{$Z_6$--II--f $SU(3)\times SU(4)^{[2]}\times SU(2)$}
There is a \non\ $Z_2$ fixed plane in the $\theta^2$ twisted sector.

$R=\left(\matrix{1&0\cr 0&1\cr}\right)\ \ K=\left(\matrix{3&0\cr 0&1\cr}\right)$

Wilson lines as stated here correspond to thes six component line as:
$A=\left(\matrix{A_3\cr A_6\cr}\right)$

For the $T$ modulus there are 2 modular groups:

\textbf{1.}\ $\Gamma^0(3)$

\ \ $A=\mathbf{0}$

\textbf{2.}$a,d=1\bmod 2$, $b=0\bmod 3$, $c=0\bmod 2$

\ \ $A={1\over 2}\left(\matrix{0&0&0&0&0&0&0&0&0&0&0&0&0&0&0&0\cr 
2&0&0&0&0&0&0&0&2&0&0&0&0&0&0&0\cr}\right)$

For the $U$ modulus there are 4 modular groups:

\textbf{1.}\ $\Gamma_0(3)$

\ \ $A=\mathbf{0}$

\textbf{2.}\ $a,d=1\bmod 2$, $c=0\bmod 6$

\ \ $A={1\over 2}\left(\matrix{0&0&0&0&0&0&0&0&0&0&0&0&0&0&0&0\cr 
2&0&0&0&0&0&0&0&2&0&0&0&0&0&0&0\cr}\right)$

\textbf{3.}\ $a,d=0\bmod 2$, $b=1\bmod 2$, $c=3\bmod 6$ and
$a,d=1\bmod 2$, $b=0\bmod 2$, $c=0\bmod 6$

\ \ $A={1\over 2}\left(\matrix{2&0&0&0&0&0&0&0&0&0&0&0&0&0&0&0\cr 
2&0&0&0&0&0&0&0&0&0&0&0&0&0&0&0\cr}\right)$

\textbf{4.}\ $a=1\bmod 2$, $b=0\bmod 2$, $c=0\bmod 3$, $d=1\bmod 2$

\ \ $A={1\over 2}\left(\matrix{2&0&0&0&0&0&0&0&2&0&0&0&0&0&0&0\cr 
0&0&0&0&0&0&0&0&0&0&0&0&0&0&0&0\cr}\right)$

\subsection{$Z_6$--II--g $SU(3)\times Sp(6)\times SU(2)$}
There is a \non\ $Z_2$ fixed plane in the $\theta^2$ twisted sector.

$R=\left(\matrix{1&0\cr 0&1\cr}\right)\ \ K=\left(\matrix{3&0\cr 0&1\cr}\right)$

Wilson lines as stated here correspond to thes six component line as:
$A=\left(\matrix{A_5\cr A_6\cr}\right)$

For the $T$ modulus there are 2 modular groups:

\textbf{1.}\ $\Gamma^0(3)$

\ \ $A=\mathbf{0}$

\textbf{2.}\ $a,d=1\bmod 2$, $b=0\bmod 3$, $c=0\bmod 2$

\ \ $A={1\over 2}\left(\matrix{0&0&0&0&0&0&0&0&0&0&0&0&0&0&0&0\cr 
2&0&0&0&0&0&0&0&2&0&0&0&0&0&0&0\cr}\right)$

For the $U$ modulus there are 4 modular groups:

\textbf{1.}\ $\Gamma_0(3)$

\ \ $A=\mathbf{0}$

\textbf{2.}\ $a,d=1\bmod 2$, $c=0\bmod 6$

\ \ $A={1\over 2}\left(\matrix{0&0&0&0&0&0&0&0&0&0&0&0&0&0&0&0\cr 
2&0&0&0&0&0&0&0&2&0&0&0&0&0&0&0\cr}\right)$

\textbf{3.}\ $a,d=0\bmod 2$, $b=1\bmod 2$, $c=3\bmod 6$ and
$a,d=1\bmod 2$, $b=0\bmod 2$, $c=0\bmod 6$

\ \ $A={1\over 2}\left(\matrix{2&0&0&0&0&0&0&0&0&0&0&0&0&0&0&0\cr 
2&0&0&0&0&0&0&0&0&0&0&0&0&0&0&0\cr}\right)$

\textbf{4.}\ $a=1\bmod 2$, $b=0\bmod 2$, $c=0\bmod 3$, $d=1\bmod 2$

\ \ $A={1\over 2}\left(\matrix{2&0&0&0&0&0&0&0&2&0&0&0&0&0&0&0\cr 
0&0&0&0&0&0&0&0&0&0&0&0&0&0&0&0\cr}\right)$

\subsection{$Z_8$--II--a $SO(10)\times SU(2)$}
There is a \non\ $Z_2$ fixed plane in the $\theta^2$ twisted sector.

$R=\left(\matrix{2&0\cr 0&1\cr}\right)\ \ K=\left(\matrix{2&0\cr 0&1\cr}\right)$

Wilson lines as stated here correspond to thes six component line as:
$A=\left(\matrix{A_4\cr A_6\cr}\right)$

For the $T$ modulus there are 6 modular groups:

\textbf{1.}\ $\Gamma^0(2)$

\ \ $A=\mathbf{0}$

\textbf{2.}\ $a,d=1\bmod 4$, $b=0\bmod 2$, $c=0\bmod 4$ and
$a,d=3\bmod 4$, $b=0\bmod 2$, $c=0\bmod 4$

\ \ $A={1\over 2}\left(\matrix{0&0&0&0&0&0&0&0&0&0&0&0&0&0&0&0\cr 
1&1&0&0&0&0&0&0&0&0&0&0&0&0&0&0\cr}\right)$

\textbf{3.}\ $a,d=1\bmod 2$, $b=0\bmod 2$, $c=0\bmod 2$

\ \ $A={1\over 2}\left(\matrix{0&0&0&0&0&0&0&0&0&0&0&0&0&0&0&0\cr 
2&0&0&0&0&0&0&0&0&0&0&0&0&0&0&0\cr}\right)$

\textbf{4.}\ $a=1\bmod 4$, $b=0\bmod 2$, $c=0\bmod 2$, $d=1\bmod 2$

\ \ $A={1\over 4}\left(\matrix{3&2&1&1&1&0&0&0&4&0&0&0&0&0&0&0\cr 
0&0&0&0&0&0&0&0&0&0&0&0&0&0&0&0\cr}\right)$

\textbf{5.}\ $a=1\bmod 4$, $b=0\bmod 2$, $c=0\bmod 8$, $d=1\bmod 4$

\ \ $A={1\over 4}\left(\matrix{3&2&1&1&1&0&0&0&4&0&0&0&0&0&0&0\cr 
0&0&0&0&0&0&0&0&0&0&0&0&0&0&0&0\cr}\right)$

\textbf{6.}\ $a=1\bmod 4$, $b=0\bmod 2$, $c=0\bmod 4$, $d=1\bmod 2$

\ \ $A={1\over 4}\left(\matrix{3&2&1&1&1&0&0&0&4&0&0&0&0&0&0&0\cr 
0&0&0&0&0&0&0&0&0&2&2&0&0&0&0&0\cr}\right)$

For the $U$ modulus there are 6 modular groups:

\textbf{1.}\ $\Gamma_0(2)$

\ \ $A=\mathbf{0}$

\textbf{2.}\ $a,d=1\bmod 4$, $c=0\bmod 8$ and
$a,d=3\bmod 4$, $c=0\bmod 8$

\ \ $A={1\over 2}\left(\matrix{0&0&0&0&0&0&0&0&0&0&0&0&0&0&0&0\cr 
1&1&0&0&0&0&0&0&0&0&0&0&0&0&0&0\cr}\right)$

\textbf{3.}\ $a,d=1\bmod 2$, $c=0\bmod 4$

\ \ $A={1\over 2}\left(\matrix{0&0&0&0&0&0&0&0&0&0&0&0&0&0&0&0\cr 
2&0&0&0&0&0&0&0&0&0&0&0&0&0&0&0\cr}\right)$

\textbf{4.}\ $a=1\bmod 2$, $b=0\bmod 2$, $c=0\bmod 2$, $d=1\bmod 4$

\ \ $A={1\over 4}\left(\matrix{3&2&1&1&1&0&0&0&4&0&0&0&0&0&0&0\cr 
0&0&0&0&0&0&0&0&0&0&0&0&0&0&0&0\cr}\right)$

\textbf{5.}\ \ $a=1\bmod 4$, $b=0\bmod 2$, $c=0\bmod 8$, $d=1\bmod 4$

\ \ $A={1\over 4}\left(\matrix{3&2&1&1&1&0&0&0&4&0&0&0&0&0&0&0\cr 
0&0&0&0&0&0&0&0&0&0&0&0&0&0&0&0\cr}\right)$

\textbf{6.}\ $a=1\bmod 2$, $b=0\bmod 2$, $c=0\bmod 4$, $d=1\bmod 4$

\ \ $A={1\over 4}\left(\matrix{3&2&1&1&1&0&0&0&4&0&0&0&0&0&0&0\cr 
0&0&0&0&0&0&0&0&0&2&2&0&0&0&0&0\cr}\right)$

\subsection{$Z_{12}$--I--a $E_6$}
There is a \non\ $Z_3$ fixed plane in the $\theta^3$ twisted sector.

$R=\left(\matrix{2&0\cr 0&2\cr}\right)\ \ K=\left(\matrix{2&0\cr 0&2\cr}\right)$

Wilson lines as stated here correspond to thes six component line as:
$A=\left(\matrix{A_1\cr A_4\cr}\right)$

For the $T$ modulus there are 2 modular groups:

\textbf{1.}\ $\Gamma^0(2)$

\ \ $A=\mathbf{0}$

\textbf{2.}\ \ $a,d=1\bmod 3$, $b=0\bmod 2$, $c=0\bmod 3$

\ \ $A={1\over 3}\left(\matrix{2&0&0&0&0&0&0&0&1&1&0&0&0&0&0&0\cr 
2&0&0&0&0&0&0&0&1&1&0&0&0&0&0&0\cr}\right)$

\section{Calculation of R and K matrices}

The matrices $R$ and $K$ allow the constraints on the modular symmetry group
of \non\ orbifolds to be expressed in a 2--dimensional form. The constraints
are obtained by ensuring that the physical winding and momentum states in the
fixed plane transform properly under the modular transformations. The physical
winding numbers and momenta in the plane must be expressable in terms of only
two independent components in each case. In particular we wish to express
the contribution to the worldsheet momentum $P$ in terms of the two independent
values of the momenta and windings, hence $K$ is defined by
\begin{eqnarray}
P=p_{soln}^Tw_{soln}=m_{\perp}^TK^Tn_{\perp}
\end{eqnarray}
where $p_{soln}$ and $w_{soln}$ are the solutions to $(Q^*)^kp=p$ and $Q^kw=w$ 
respectively and are six dimensional vectors and $m_{\perp},n_{\perp}$ are the
corresponding two dimensional momentum and winding numbers. An example of the
calculation of $K$ for a specific orbifold was given in appendix C.
Similarly, the $R$ matrix is introduced when expressing the constraints in
the presence of non--zero Wilson lines in terms of only two independent lines.
The Wilson lines involved in the problem are those which correspond to the
windings of $w_{soln}$. For example, in the case of the $Z_6$--II--f of appendix
C, $w_{soln}$ is given by $w_{soln}=(n_1,2n_1,-n_1,0,0,n_6)^T$, and so the
Wilson lines involved in the problem are the components $A_1,\,A_2,\,A_3$
associated with windings in one direction in the fixed plane and $A_6$ 
associated with the windings in the other direction. It might appear that there 
are then in fact 4 different Wilson line components involved in the problem,
but the number of independent Wilson line components is reduced to only two
in all problems by the point group invariance condition, which in this case
yields the relations $A_2\approx A_3,\,A_1\approx 2A_2\approx 0$. So in the case
of the $Z_6$--II--f orbifold the Wilson line contribution can be expressed in
terms of the independent components $A_2$ and $A_6$. The $R$ matrix is then
defined by
\begin{eqnarray}
Aw_{soln}=A_{\perp}Rn_{\perp}
\end{eqnarray}
where $A$ is the six dimensional Wilson line and $A_{\perp}$ is the two 
dimensional line consisting of the two independent  components of the
six dimensional line which contribute, as discussed above, and $n_{\perp}$
is as defined above.

%references

\end{document}